\newcommand\aap{{A\&A}}
\newcommand\aaps{{A\&AS}}
\newcommand\aj{{AJ}}
\newcommand\apj{{ApJ}}
\newcommand\apjl{{ApJ}}
\newcommand\apjs{{ApJS}}
\newcommand\araa{{ARA\&A}}

\newcommand\jaa{{JA\&A}}

\newcommand\mnras{{MNRAS}}
\newcommand\nat{{Nat}}

\newcommand\pasp{{PASP}}
\documentclass[useAMS,usenatbib]{mn2e}

\usepackage{epsfig}
\usepackage{rotating}
\usepackage{amssymb}
\usepackage{multirow}
\usepackage{url}

\title[Is there really a dichotomy in AGN jet power?]{Is there really a dichotomy in AGN jet power?}
\author[J. W. Broderick and R. P. Fender]{J. W. Broderick\thanks{E-mail:
J.Broderick@soton.ac.uk} and R. P. Fender\\
School of Physics and Astronomy, University of Southampton, Highfield, Southampton SO17 1BJ}
\begin{document}


\pagerange{\pageref{firstpage}--\pageref{lastpage}} \pubyear{2011}

\maketitle

\label{firstpage}

\begin{abstract}

To gain new insights into the radio-loud/radio-quiet dichotomy reported for active galactic nuclei, we examine radio loudness as a function of Eddington ratio for a
previously published sample of 199 AGN from five different populations. After initially considering radio loudnesses derived using total radio luminosities, we repeat the investigation using {\em core} radio luminosities only, applying a previously established mass correction for
these core luminosities. In both cases, for Eddington ratios $< 1$
per cent, Fanaroff--Riley type I and broad-line radio galaxies are on
average more radio-loud than Seyfert and low-ionization nuclear
emission-line region galaxies. However, the distribution of radio
loudnesses for the mass-corrected, core-only sample is much narrower
than that of the clearly bimodal total radio loudness
distribution. The advantages and disadvantages of using core- or
lobe-dominated radio luminosity as a measure of instantaneous jet
power are discussed.  We furthermore compare the core and total radio
luminosities for the entire sample, as well as illustrating the
importance of the mass term by comparing the AGN with a sample of black hole
X-ray binaries. We conclude that if the mass-corrected core radio
luminosity is a good measure of jet power, then black hole spin may have considerably
less impact on jet power than previously reported, or that our sample
does not include the extremes of spin. If the spread in jet power is 
small then we suggest that characteristics of the ambient environment and/or the radio source age
could be equally as important in producing a radio-loud/radio-quiet dichotomy seen in total radio
luminosity.
 
\end{abstract}


\section{Introduction}\label{introduction}

The `radio loudness' of an active galactic nucleus (AGN) is usually defined as the ratio of its radio and optical flux densities or luminosities at two specific frequencies. A possible bimodality in the radio loudness distribution of the AGN population, the so-called `radio-loud/radio-quiet dichotomy', is an often-debated, somewhat contentious, and ongoing topic of study \citep*[e.g.][]{strittmatter80,sramek80,condon81,kellermann89,miller90,miller93,xu99,white00,ivezic02,cirasuolo03a,cirasuolo03b,gopal08,zamfir08,singal11}. The resolution of this issue is crucial if fundamental questions relating to the physics of black hole (BH) accretion, jet formation and feedback are to be fully addressed.

One standard definition of radio loudness that has been used particularly in quasar studies is the ratio $S_{\nu_{5}}/S_{\nu_{B}}$, where $S_{\nu_{5}}$ and $S_{\nu_{B}}$ are the monochromatic 5 GHz radio and nuclear 4400 \AA\ ($B$-band) flux densities, respectively \citep[][]{kellermann89}. Radio-loud quasars were at first considered to be those with $S_{\nu_{5}}/S_{\nu_{B}} \gtrsim 10$, while for most radio-quiet quasars $0.1 < S_{\nu_{5}}/S_{\nu_{B}} < 1$  \citep[e.g.][]{peterson97}. However, in more recent times, BH mass measurements\footnote{In this paper we use the dimensionless BH mass $M~\equiv~M_{\rm BH}/M_{\sun}$.} have allowed a more sophisticated approach to be adopted when studying the radio loudnesses of AGN: the dependence of radio loudness on the Eddington ratio, $\lambda$ ($\equiv L_{\rm bol}/L_{\rm Edd}$, where $L_{\rm bol}$ is the nuclear radiative bolometric luminosity and $L_{\rm Edd}$ is the Eddington luminosity). For example, \citet*[][]{sikora07}, henceforth referred to as SSL07, investigated the radio loudnesses of a total of 199 sources spread across five different populations: broad-line radio galaxies (BLRGs), radio-loud quasars (RLQs), Seyfert and low-ionization nuclear emission-line region galaxies (SGs and LINERs), Fanaroff--Riley type I radio galaxies (FR I RGs) and Palomar--Green quasars (PGQs). In particular, SSL07 showed that there are two distinct, approximately parallel tracks when log($R$) is plotted as a function of log($\lambda$); in this case the radio loudness parameter $R$ was defined as

\begin{equation}\label{eqn_radio_loudness}
R \equiv L_{\nu_{5}}/L_{\nu_{B}},
\end{equation}

where $L_{\nu_{5}}$ and $L_{\nu_{B}}$ are the 5 GHz and nuclear 4400 \AA\ monochromatic luminosities, respectively. Along both tracks, log($R$) increases as log($\lambda$) decreases (also see \citealt[][]{ho02}, who found a similar trend), a result analogous to that previously established for BH X-ray binaries (XRBs) in the `low/hard' state \citep*[e.g.][]{gallo03} . The upper, `radio-loud sequence' is occupied by BLRGs, RLQs and FR I RGs, while the lower, `radio-quiet' sequence comprises SGs, LINERs and PGQs. The bimodality is most apparent at log($\lambda) \lesssim -2$ ($\lesssim 1$ per cent Eddington).

SSL07 interpreted their findings in terms of BH spin \citep[e.g.][]{blandford77,mckinney05}; powerful jets can be produced through the extraction of the rotational energy of the BH by an external magnetic field that is supported by the accretion disc. Giant ellipticals on the radio-loud track are postulated to host high-spin BHs, while low-spin BHs in spiral/disc galaxies are found on the radio-quiet track \citep*[also see][]{volonteri07,sikora09,lagos09,tchekhovskoy10,mcnamara11,daly11,sansigre11}. Hypotheses can then be drawn regarding the relationship between spin evolution and galaxy merger events (e.g. discussion in SSL07). At Eddington ratios above $\sim$1 per cent, where there is some mixing of the tracks, there is also thought to be an additional dependence on accretion state, analogous to what is seen in BH XRBs, where in disc-dominated `high/soft' states the jet is strongly suppressed \citep*[e.g.][]{maccarone03,fender04}.

In their analysis, SSL07 used the total monochromatic radio power at 5 GHz, that is core plus extended emission, when calculating $R$. An alternative method would be to only consider the radio luminosity from the core. Currently there is no general consensus in the literature about which method is best; there are arguments for and against each approach. \citet{white00} did not find a gap in the radio loudness distribution of sources in the Faint Images of the Radio Sky at Twenty Centimetres \citep*[FIRST;][]{becker95} Bright Quasar Survey (FBQS); the 1.4 GHz FIRST survey has an angular resolution of 5 arcsec, which may result in a loss of sensitivity to extended emission \citep[e.g. see discussion in][]{laor03}. Though FIRST may generally underestimate the flux density of the extended component, \citet*[][]{rafter11} found that once this emission is taken into account, there is a radio loudness bimodality in their sample of AGN from the Sloan Digital Sky Survey \citep[SDSS;][]{york00}: those sources extended in FIRST are more radio-loud on average than core-only sources. Similarly, a link between radio loudness bimodality and the use of total, rather than core, radio luminosities was suggested by the work of \citet[][]{terashima03}. Given the results of these various investigations, one can therefore speculate that the two tracks in SSL07 may approach each other or even merge if total radio luminosities are replaced with core radio luminosities.

Two other studies which only considered the radio emission from the core were by \citet*[][]{merloni03} and \citet*[][]{falcke04}, who constructed `fundamental planes of BH activity' using samples of both AGN and BH XRBs. While they too did not find significant evidence for a radio-loud/radio-quiet dichotomy, a dependence between core radio luminosity, core X-ray luminosity and BH mass was observationally established. In \citet[][]{merloni03}, the relationship is given as

\begin{equation}\label{eqn_merloni}
L_{5, \rm \: core} \propto L_{X}^{0.6} M^{0.8},
\end{equation}

where $L_{5, \rm \: core}$, the 5 GHz core luminosity ($\equiv \nu_{5}L_{\nu_{5, \rm \: core}}$ where $\nu_{5} = 5$ GHz) and $L_{X}$, the 2--10 keV nuclear X-ray luminosity are both measured in erg s$^{-1}$. Note that very similar conclusions were reached in \citet[][]{falcke04}; see also \citet*[][]{koerding06} and \citet[][]{gultekin09}. These studies confirmed the earlier theoretical prediction of \citet[][]{heinz03}, who derived a model-independent relation for scale-invariant jets \citep[i.e. in the inner regions of the jet only; also see further discussion in][]{merloni03}. The mass term in equation~\ref{eqn_merloni} arises from the fact that the jet volume varies faster than the mass, and thus jets from more massive BHs are more radiatively efficient at a given accretion rate as a result of the decrease in optical depth. \citet[][]{heinz03} noted that random variations in the BH spin parameter could simply introduce an intrinsic scatter to the fundamental plane relation that may be independent of $M$. Therefore, correcting $R$ for the effects of the mass term should give a clearer picture of the contribution of BH spin to any radio loudness dichotomy, and in turn the distribution of BH spin in AGN.

In this paper, we revisit the study of SSL07 to determine how the use of core radio luminosities, and the application of a mass correction term affects the bimodality observed in the log($R$)--log($\lambda$) plane. In Section~\ref{AGN samples}, we give an overview of how core luminosities were obtained and mass corrections derived for the sample from SSL07, before reexamining the tracks in the log($R$)--log($\lambda$) plane in Section~\ref{sikora revisit}. We discuss our new findings in Section~\ref{discussion}, before concluding in Section~\ref{conclusions}.

We assume the same $\Lambda$CDM cosmological model as in SSL07, that is $\Omega_{\scriptsize{\textnormal{M}}} = 0.3$, $\Omega_{\Lambda} = 0.7$, and $H_{0} = 70$ km s$^{-1}$ Mpc$^{-1}$. We also define $R$ as in equation~\ref{eqn_radio_loudness}, but with $L_{\nu_{5}}$ calculated from the core only. Furthermore, as in SSL07, we assume that $L_{\rm bol} = 10\nu_{B}L_{\nu_{B}} = 10 L_{B}$. All values of $\lambda$ have been calculated using the $L_{B}$ values and BH masses presented in SSL07.
\newline

\section{The SSL07 AGN sample: core radio luminosities and mass corrections}\label{AGN samples}

\begin{table*}
\caption{Median values of various source properties. Some values are more uncertain than others due to the presence of limits. The medians have been calculated using the data from SSL07, except for the core radio luminosity column (this paper).}
\label{table: general properties}
\begin{center}
\begin{tabular}{cccccccc}
\hline
&     &  \multicolumn{6}{c}{Medians} \\
\cline{3-8} 
Subsample & $N$ & $z$ & $\log{(L_{B})}$ & $\log{(L_{5, \rm \: total})}$ & $\log{(L_{5, \rm \: core})}$ & $\log{(\lambda)}$ & $\log{(M)}$ \\
 & & & (erg s$^{-1}$) & (erg s$^{-1}$) & (erg s$^{-1}$) & &  \\
\hline
BLRGs & $37$ & $0.157$ & $44.1$ & $42.3$ & $\sim$$41.1$ &  $-1.6$ & $8.4$  \\
RLQs & $50$ & $0.325$ & $45.1$ & $42.8$ & $\sim$$42.3$ &  $-0.8$ & $8.8$ \\
SGs and LINERs & $38$ & $0.009$ & $42.3$ & $38.4$ & $\sim$$37.7$ & $-2.1$ & $7.6$ \\ 
FR I RGs & $31$ & $0.025$ & $\sim$$41.2$ & $41.1$ & $39.8$ & $\sim$$-4.8$ & $8.6$ \\
PGQs & $43$ & $0.144$ & $45.0$ & $39.4$ & $\sim$$39.1$ & $-0.3$ & $8.2$ \\
\hline
\end{tabular}
\end{center}
\end{table*}

As mentioned in Section~\ref{introduction}, SSL07 used a total of 199 AGN in their study, spread across five different populations: BLRGs (37 sources), RLQs (50 sources), SGs and LINERs (38 sources, three of which are LINERs), FR I RGs (31 sources) and PGQs (43 sources).\footnote{There are 200 entries in the SSL07 data tables, but Perseus A has been counted twice: firstly as NGC 1275 in the SG/LINER subsample, and secondly as 3C 84 in the FR I RG subsample. Slightly different measurements are listed in the two separate data table entries; we note especially that the BH mass changes by 0.6 dex. We chose to include this source in the FR I RG subsample only (see Section 3.1 in SSL07 for a discussion on how Perseus A is an outlier when considered as part of the SG/LINER subsample), and we have used the data in the FR I RG table in SSL07 (table 4).}\label{footnote perseus a} Median values for a number of source properties for each subsample are shown in Table~\ref{table: general properties}. In addition, radio {\em core} luminosities for the sources in the SSL07 sample are presented in Appendix~\ref{appendix radio fluxes} (Tables \ref{table: radio cores1}--\ref{table: radio cores5}); we now briefly summarize how these values were obtained.

The 5 GHz monochromatic core luminosity was calculated using the standard formula

\begin{equation}\label{eqn_core_luminosity}
L_{\nu_{5, \rm \: core}} = 4 \pi D_{\rm L}^2 S_{\nu_{\rm 5, \: core}} (1+z)^{-(1+\alpha)},
\end{equation}
where $D_{\rm L}$ is the luminosity distance for our assumed cosmology, $S_{\nu_{\rm 5, \: core}}$ is the 5 GHz core flux density, $z$ is the redshift and $\alpha$ is the radio spectral index (defined using the convention $S_{\nu} \propto \nu^{\alpha}$). If there were multiple flux density measurements for a given source, then these were averaged together. In some cases, only a luminosity was available in a particular reference in the literature, and a correction was applied so that it was valid for our cosmological model. Unless radio spectral index information was available, we assumed a spectral index of 0 when calculating the core luminosity. This will introduce some scatter into our $R$ measurements. However, because all of the sources in the sample are at $z < 0.5$, with a median redshift of $0.138$, we would expect that in most cases the uncertainty introduced into log($R$) should be small ($ \lesssim \pm 0.1$). In other words, uncertainties in the radio $k$-correction term in equation~\ref{eqn_core_luminosity} do not significantly affect our investigation.

Core measurements at 5 GHz for the BLRG, RLQ and FRI RG subsamples were obtained from a variety of references in the literature. For the SGs and LINERs, SSL07 used the total radio luminosities from \citet[][]{ho01} and \citet[][]{ho02}; both these papers also include `nuclear' radio luminosities, but they were computed using the emission from all of the components that are thought to be associated with the active nucleus. We instead searched for core-only flux densities in the literature. The core radio luminosities were then calculated using the distance measurements given in \citet[][]{ho01} and \citet[][]{ho02}; as in SSL07, we corrected distances greater than 40 Mpc to take into account that our assumed cosmological model differs from those used in \citet[][]{ho01} and \citet[][]{ho02}. Lastly, core luminosities for the PGQs were derived from the flux densities presented in \citet[][]{kellermann89} and \citet[][]{miller93}. \citet{miller93} reanalysed the 5 GHz data used in the \citet[][]{kellermann89} study, and if a core flux density for a particular source was available from both studies, we chose to average the measurements together.

For some of the more compact sources, we used either the integrated or peak 5 GHz flux density of the total radio source. Nonetheless, the corresponding observations were usually carried out at sufficiently high angular resolution ($\lesssim 10$ arcsec), and in the majority of cases
no source extension or only partial extension was reported. Flux densities from ATCA and VLA calibrator tables were also used where available. Therefore, these measurements should provide reasonable estimates for the core luminosities. Also, it was sometimes necessary to extrapolate from or interpolate between measurements at neighbouring frequencies, either using known spectral index information or assuming that $\alpha=0$. For 34 of the sources, we were only able to obtain an upper limit for the core luminosity (17 per cent of the total sample of 199 AGN). See Appendix~\ref{appendix radio fluxes} for further details.

The mass correction term can be deduced as follows. Rearranging equation~\ref{eqn_merloni} gives

\begin{equation}\label{eqn_merloni2}
\frac{L_{5, \rm \: core}}{L_{X}} \propto \left(\frac{L_{X}}{M}\right)^{-0.4}M^{0.4}.
\end{equation}
If the nuclear optical and X-ray luminosities only differ by some constant factor (an assumption which we discuss further in Section~\ref{discussion}), then we can replace $L_{X}$ with $L_{B}$ in equation~\ref{eqn_merloni2}. Furthermore, given that $L_{B} \equiv \nu_{B}L_{\nu_{B}}$, $L_{5, \rm \: core} \equiv \nu_{5}L_{\nu_{5, \rm \: core}}$, our assumption of $L_{\rm bol} = 10 L_{B}$ and the definition of the Eddington luminosity itself, it follows that

\begin{equation}
R \propto \lambda^{-0.4}M^{0.4},
\end{equation}
or
\begin{equation}\label{eqn_merloni4}
\log(R) = -0.4\log(\lambda) + 0.4 \log(M) + \rm \: constant.
\end{equation}
When plotted in the log($R$)--log($\lambda$) plane, the mass-corrected radio loudness is therefore of the form log($R)- 0.4 \log(M)$. The mass terms were calculated using the BH mass data from SSL07. SSL07 concluded that their results are not affected significantly by errors in the BH mass values, which were obtained using a variety of methods. We estimate that the typical uncertainty should be $\lesssim$ 0.5 dex \citep[e.g. see][]{mclure02,vestergaard02,woo02}. 

In addition, equation~\ref{eqn_merloni4} can be related to the total jet power, $P_{\rm J}$. For flat-spectrum emission from the radio core,  

\begin{equation}\label{eqn_blandford}
L_{5, \rm \: core} \propto P_{\rm J}^{17/12}
\end{equation}     
\citep[][]{blandford79}; empirical evidence for this relationship was found by \citet*[][]{koerding06b}. Thus, at a given value of log$(\lambda)$, a 1 dex difference in $R$ corresponds to a difference in jet power of $\frac{12}{17} \approx 0.7$ dex.     

\section[]{Reanalysing the distribution of points in the log($R$)--log($\lambda$) plane}\label{sikora revisit}

\begin{figure}
\epsfig{file=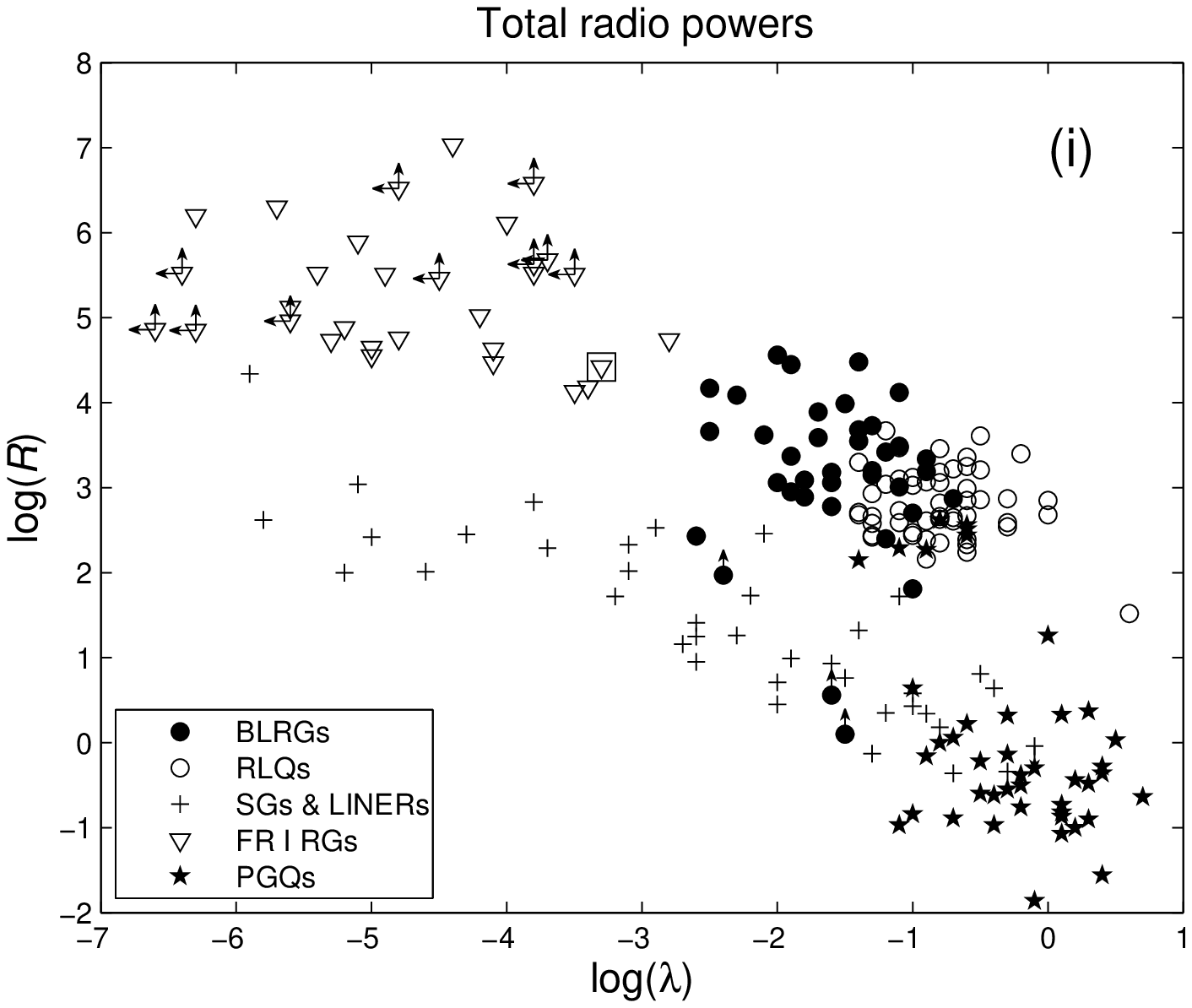,width=8.40cm}\newline
\epsfig{file=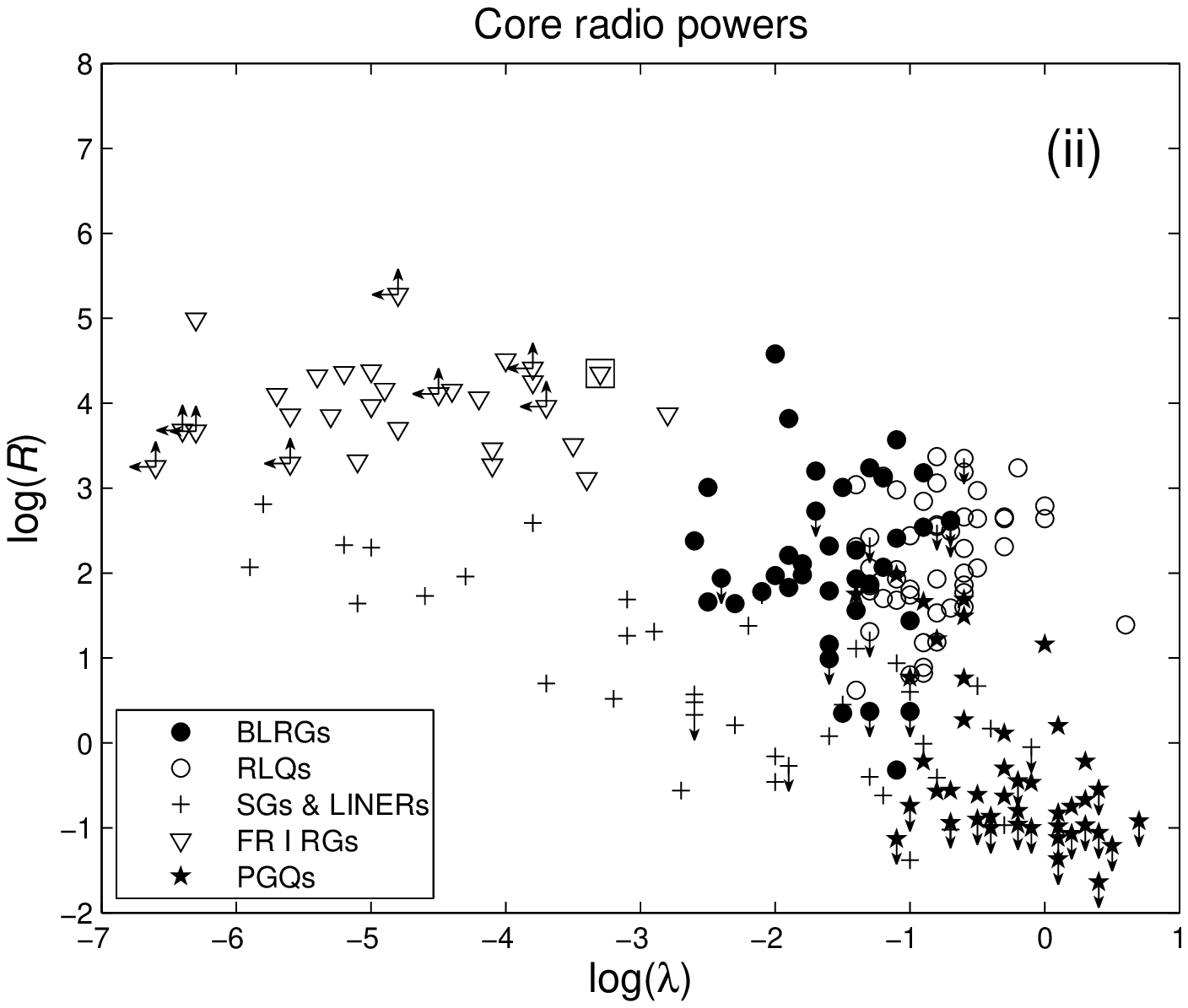,width=8.40cm}\newline
\epsfig{file=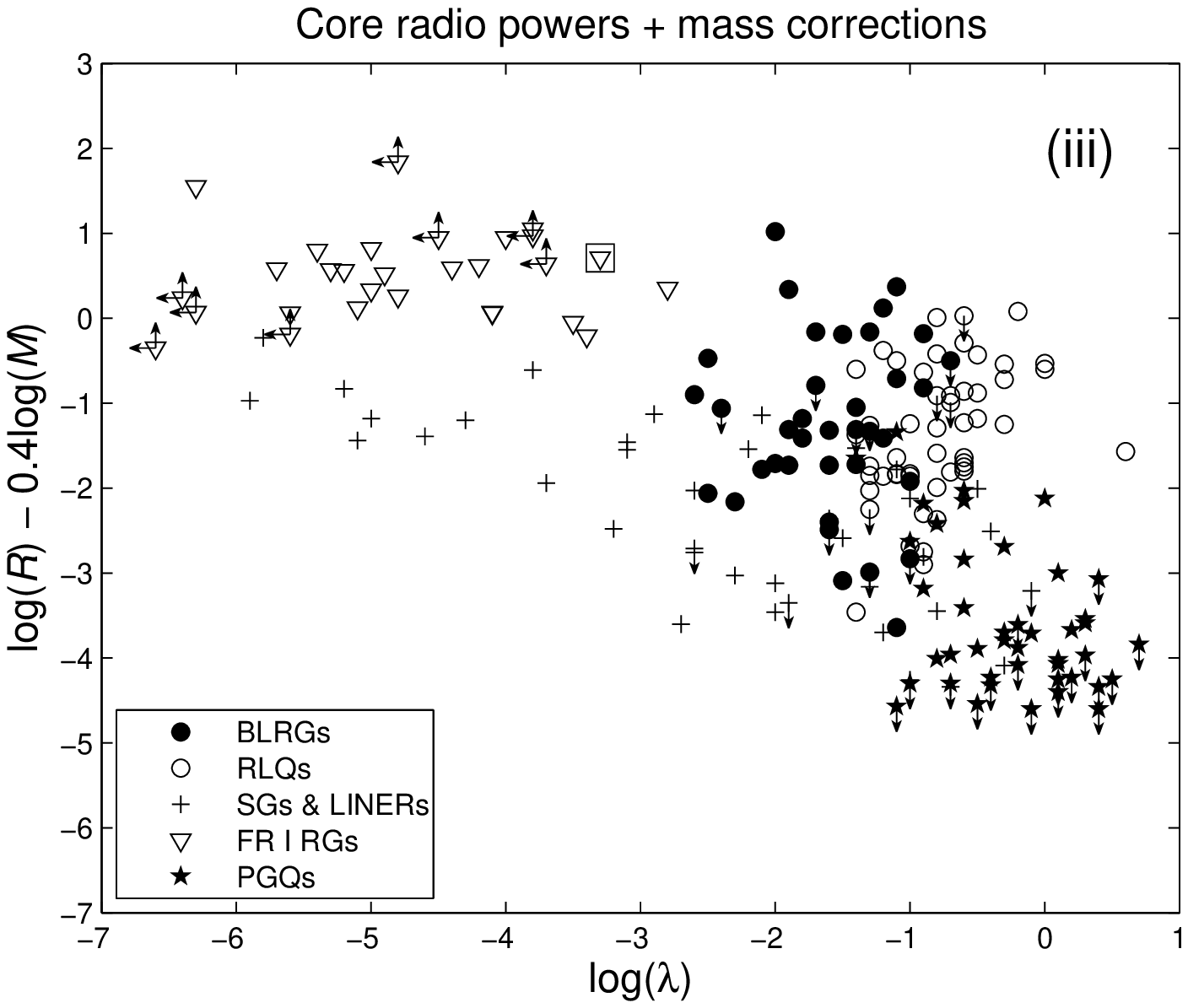,width=8.40cm}
\caption{$\log(R)$ plotted as a function of $\log(\lambda)$ for three different scenarios: (i) original data from SSL07 (with minor modifications); (ii) radio loudnesses calculated using core luminosities; and (iii) radio loudnesses calculated using core luminosities, with a mass correction term included on the vertical axis. A square has been drawn around the data point for Perseus A (NGC 1275 / 3C 84) in all three panels (see footnote~\ref{footnote perseus a}).}
\label{fig:plot1}
\end{figure}

Figure~\ref{fig:plot1} demonstrates how the distribution of points in the log($R$)--log($\lambda$) plane changes depending on whether total or core radio luminosities are used, and if a mass correction term is applied. Firstly, we show the original data from SSL07 in panel (i), though with some slight modifications\footnote{In addition to the changes for Perseus A outlined in footnote~\ref{footnote perseus a}, we have corrected a minor discrepancy in SSL07 regarding the radio properties of Mrk 590 in the SGs and LINERs subsample.}; the distribution is clearly bimodal. In panel (ii), radio loudnesses derived from core radio luminosities have been plotted; though $R$ still increases with decreasing $\lambda$, it is immediately apparent that the gap between the upper and lower tracks has closed significantly. However, it is the use of {\em both} core luminosities and mass corrections that results in the greatest convergence of the two sequences (panel iii).

For the three above scenarios, we attempted to quantify any possible differences between the distributions of RGs and SGs/LINERs. In the analysis that follows in this section, we first applied a log($\lambda) < -2$ cutoff to minimize the effect of varying X-ray states and their coupling to jet production at higher Eddington ratios \citep[][]{fender04}. Secondly, the four FR I RGs with log($\lambda$) $< -6$ (three of which have limits in $R$ and $\lambda$) were excluded so that similar baselines in log($\lambda$) were covered by both subsets. Nonetheless, the analysis is complicated by the fact that a number of remaining sources have limits in $R$ and/or $\lambda$, especially in the FR I RG subsample. We generally have treated any limits as detections; the potentially large uncertainties associated with a range of datapoints in the log($R$)--log($\lambda$) plane are discussed in more detail in Section~\ref{discussion}. The FR I RG lower $R$ limits imply that the average differences in radio loudness between the tracks may be somewhat larger than is suggested here, but because these lower $R$ limits result from upper $L_{B}$ limits (see table 4 in SSL07), the gap between the sequences would still be smallest for case (iii) in Figure~\ref{fig:plot1}.  

The two-dimensional Kolmogorov--Smirnov (KS) test\footnote{We used the \citet{fasano87} algorithm; C code can be found in \citet[][]{press92}. An additional bug fix for this code is described at \url{http://www.nr.com/forum/showthread.php?t=576}.} was used to evaluate whether the distributions of RGs and SGs/LINERs are drawn from different parent populations; the $P$-values are shown in Table~\ref{table: KS tests}. Unsurprisingly, there is very strong evidence that the RG and SG/LINER subsets are drawn from different parent populations in panel (i) of Figure~\ref{fig:plot1}; apart from a handful of sources, RGs are found on the upper sequence, and SGs/LINERs on the lower sequence. Though the bimodality is not as pronounced in the other two cases, the $P$-values are still highly significant. The very small $P$-values are mainly due to the differences in log($R$): there is no significant evidence from one-dimensional KS tests that the log($\lambda$) distributions of the RGs and SGs/LINERs are drawn from different parent populations (but note that an intrinsic dependence on $\lambda$ could exist given the upper $\lambda$ limits for some of the FR I RGs). 

\begin{table}
\caption{Two-dimensional KS test statistics for $-6 <$ log$(\lambda) < -2$. The number of sources included from each subsample is stated. When core radio luminosities are used, the number of RGs drops to 31 because there are insufficient data for two of the FR I RGs (see Table~\ref{table: radio cores4}).}
\label{table: KS tests}
\begin{tabular}{ccccc}
\hline
\multicolumn{1}{c}{Scenario} & \multicolumn{1}{c}{$N$} & \multicolumn{1}{c}{$N$} & \multicolumn{1}{c}{$P$-value}  \\
& \multicolumn{1}{c}{(RGs)} & \multicolumn{1}{c}{(SGs/LINERs)} \\
\hline
$L_{5, \rm \: total}$ & $33$ & $20$ & $4 \times 10^{-8}$   \\
$L_{5, \rm \: core}$ & $31$ & $20$ & $5 \times 10^{-7}$  \\
$L_{5, \rm \: core}$ & $31$ & $20$ & $4 \times 10^{-6}$ \\
$+$ mass correction &       &      &                      \\
\hline
\end{tabular}
\end{table}

\begin{figure}
\epsfig{file=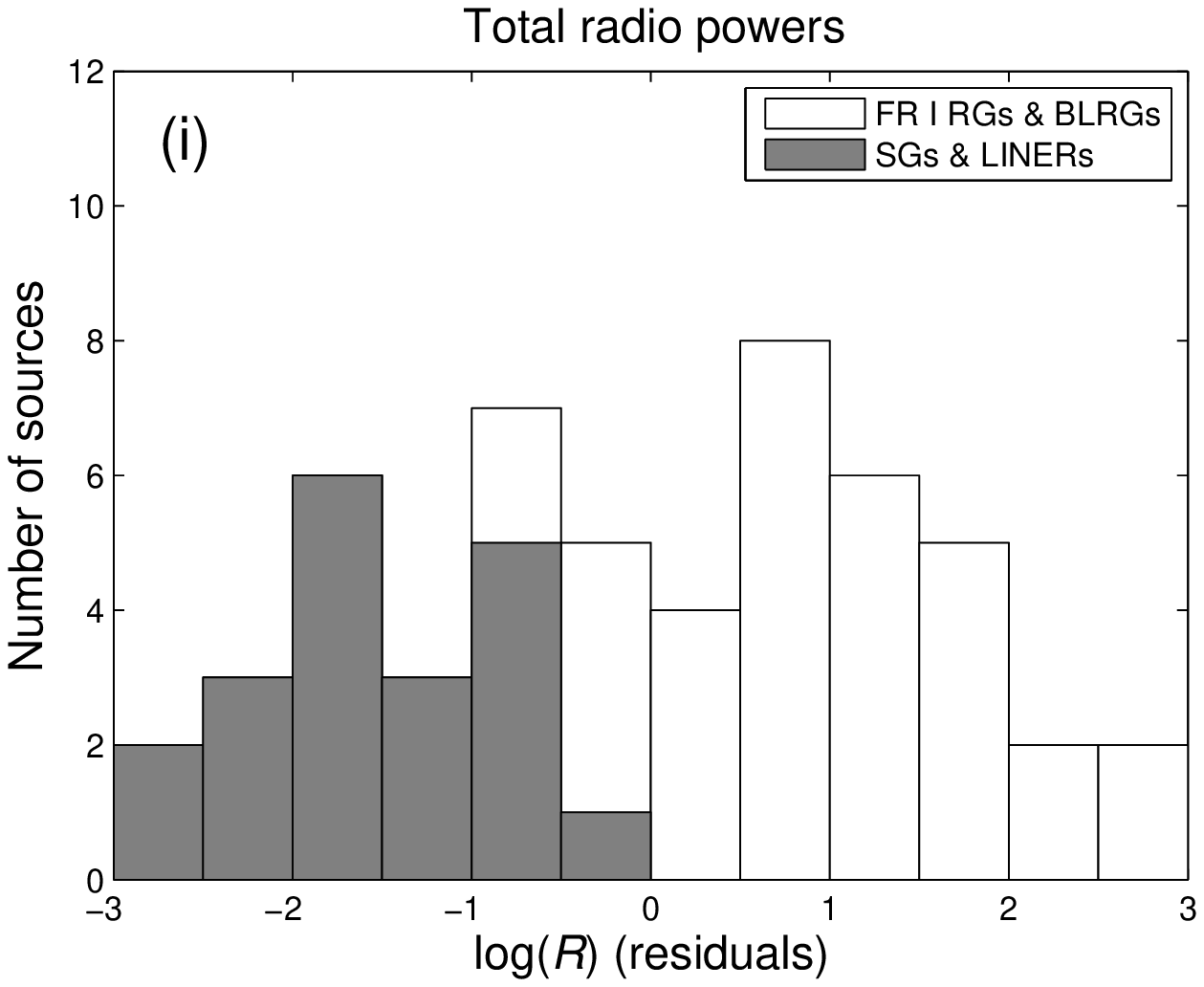,width=8.25cm}\newline
\newline
\epsfig{file=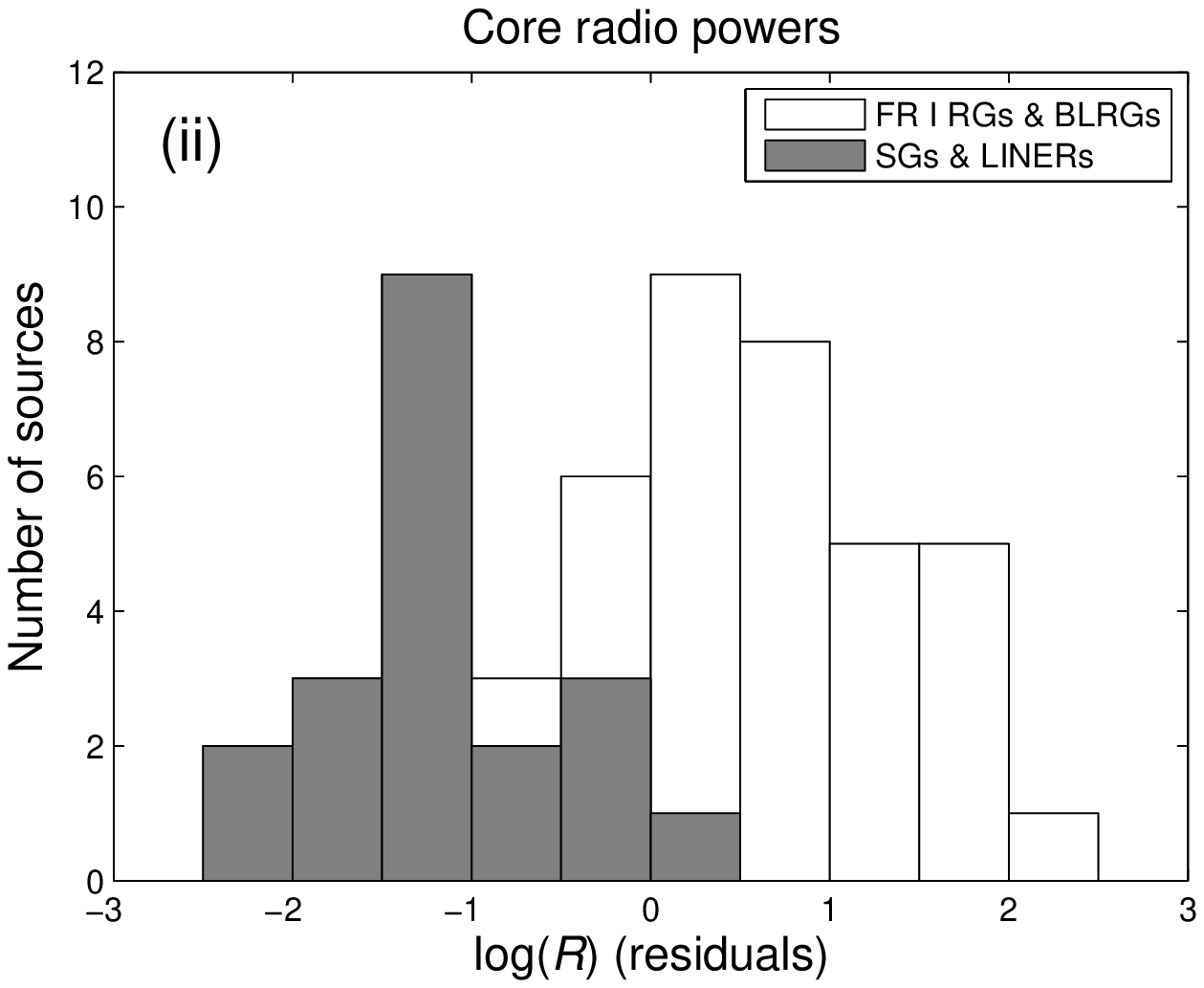,width=8.25cm}\newline
\newline
\epsfig{file=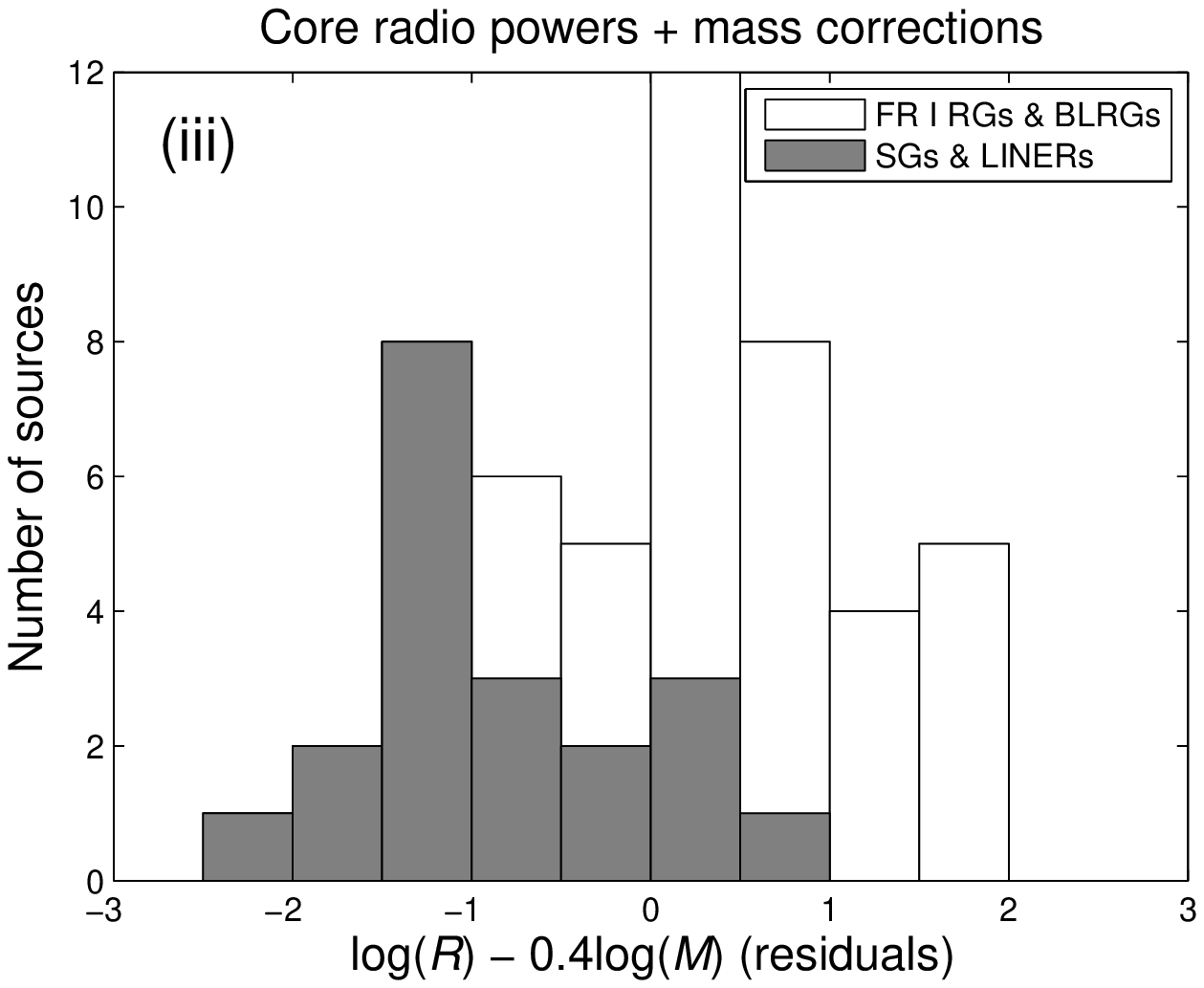,width=8.25cm}
\caption{Stacked histograms of the radio loudness residuals in the range $-6 < \log(\lambda) < -2$ .}
\label{fig:plot2}
\end{figure}

One can also model the dependence of log($R$) on log($\lambda$) by fitting a linear function (in log-log space) to each of the RG and SG/LINER subsets in all three panels of Figure~\ref{fig:plot1} for $-6 < \log({\lambda}) < -2$. Standard linear regression was found to give similar fit coefficients (within the uncertainties) to those obtained from the \textsc{schmittbin} task in the \textsc{iraf/stsdas} package, which takes into account data with limits using the method of \citet[][]{schmitt85}. The slopes of the various lines range from $\sim$$-0.6$ to $-0.4$, broadly consistent with equation~\ref{eqn_merloni4}. The average vertical offsets between the lines are about 2.6 dex (panel i), 2.1 dex (panel ii) and 1.5 dex (panel iii). 

Lastly, we also fitted linear functions (again in log-log space) through the combined distribution of RGs and SGs/LINERs with $-6 < \log(\lambda) < -2$ in each panel of Figure~\ref{fig:plot1}. The fits from standard linear regression are $\log(R) = -0.71 \log(\lambda) + 1.1$ (panel i), $\log(R) = -0.69 \log(\lambda) + 0.05$ (panel ii), and $\log(R)= -0.61 \log(\lambda) - 2.95$ (panel iii). Again, similar coefficients are found if the data with limits are treated more rigorously. Stacked histograms of the radio loudness residuals (i.e. the vertical offsets from the line of best fit) are shown in Figure~\ref{fig:plot2}. The residuals for the RGs and SGs/LINERs overlap only slightly in panel (i). For the unbinned data, the difference between the median residual of the RG subset and the median residual of the SG/LINER subset is 2.5 dex; the range covered by both subsets together is 5.6 dex. For the core radio powers (panel ii), the difference between the medians has decreased to 2.0 dex and the range to 4.5 dex. The implementation of the mass correction (panel iii) further reduces the difference between the medians (to 1.7 dex), as well as the range (to 4.1 dex). In the third case, the standard deviation of the unbinned distribution is 1.0 dex. Despite the changes in the histogram properties, there is no evidence from one-dimensional KS tests (on the unbinned data) that the three sets of combined RG $+$ SG/LINER residuals are drawn from different parent populations. A larger sample size is needed for log($\lambda$) $< -2$. 

\begin{table*}
\caption{Advantages and disadvantages of using core or total radio luminosities in studies of the radio loudness properties of AGN.}
\label{table: pros cons}
\begin{center}
\begin{tabular}{cll}
\hline
& \multicolumn{1}{c}{Core radio luminosities} & \multicolumn{1}{c}{Total radio luminosities} \\
\hline
Advantages & * Probing a similar spatial region to  & * Emission from lobes is isotropic. \\
           &  optical/X-ray observations of the nucleus.  & * More global picture of jet power.                             \\
      & * Tracing a similar history to  &                               \\
      & optical/X-ray observations of the nucleus. &                               \\
      & &                               \\
Disadvantages & * Orientation effects: beaming.  & * Extended flux can be affected significantly by    \\
   & * Core flux can be significantly variable. &   the source age or environment.                             \\
   & * Very high-resolution observations  &   * Timescales and spatial scales of the extended                             \\
   & essential in many cases. & emission not well-aligned with the corresponding                                \\
   & * Possible issues concerning  &   scales for the nuclear optical/X-ray emission.                          \\
   &   misidentification of the core if there &  * May need sufficient short baselines to record all                          \\
   &   are several radio components.&     of the extended flux.                                  \\
      & &                               \\
\hline
\end{tabular}
\end{center}
\end{table*}

Thus, in summary, replacing total radio luminosities with core radio luminosities, and additionally applying a mass correction, narrows the gap in radio loudness between the upper and lower sequences in the log($R$)--log($\lambda$) plane by about an order of magnitude in the range $-6 < \log(\lambda) < -2$. RGs are on average more radio-loud than the SGs and LINERs by $\sim$1.6 dex, which is equivalent to a difference of about an order of magnitude in jet power (equation~\ref{eqn_blandford}).

\section{Discussion}\label{discussion}

\subsection{Core versus total radio luminosities}\label{core_vs_lobe}

We have demonstrated that the significance and width of the radio loudness bimodality in the log($R$)--log($\lambda$) plane is closely linked to the measure of radio power adopted, and its dependence on BH mass. Table~\ref{table: pros cons} summarises the advantages and disadvantages of using core or total radio luminosity as a measure of instantaneous jet power. The primary argument for not using core luminosities is that they are susceptible to relativistic beaming effects \citep[e.g. in the context of $R$ bimodality by][]{laor03}, depending on the orientation of the radio source with respect to the line of sight; we discuss this further in Section~\ref{section beaming}. On the other hand, for example, recent or renewed activity in the nucleus seen, say, in the optical or X-ray bands will not be related directly to the extended radio emission from a jet that may have propagated hundreds of kpc over timescales of $\sim$$10^{7}$ yr. In this situation, the total integrated flux density is an average measure of the jet power over a much longer period of time. Therefore, studies that compare total radio luminosities with nuclear optical or X-ray luminosities may be very misleading, especially if the sources are lobe-dominated. By restricting the analysis to the nuclear region for all wavebands, we ensure that similar spatial scales, and importantly, similar timescales, are probed.

\begin{figure}
\epsfig{file=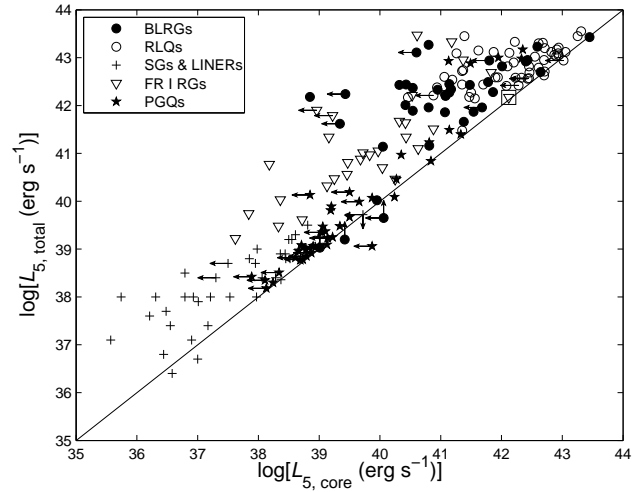,width=8.6cm}
\caption{Total versus core 5 GHz radio luminosities for the sample from SSL07. A line of equality is shown; points below the line without limits suggest variable sources. A square has been drawn around the data point for Perseus A (NGC 1275 / 3C 84; see footnote~\ref{footnote perseus a}).}
\label{fig:plot3}
\end{figure}

In Figure~\ref{fig:plot3}, we compare the core radio powers presented in this paper with the total radio powers from SSL07. Core and total 5 GHz radio power are observed to be correlated; similar correlations between core and total radio luminosity have been seen in other studies \citep[e.g.][]{giovannini88,giovannini01}. The BLRGs and FR I RGs are typically dominated by an extended component: the median core-to-total radio luminosity ratios are $\sim$7 per cent (BLRGs) and $\sim$6 per cent (FR I RGs). For the RLQs, as well as the SGs and LINERs, the majority of the radio power also originates from an extended component, but not to the same extent as for the BLRGs and FR I RGs: the median core-to-total ratios are $\sim$25 per cent (RLQs) and $\sim$20 per cent (SGs and LINERs). The PGQs may have the largest core fraction on average, but given that only core upper limits were obtained for 40 per cent of this subsample, it is not possible to draw a firm conclusion at this stage (median ratio $\gtrsim 20$ per cent).

\subsection{Possible effects of relativistic beaming}\label{section beaming} 

Could relativistic beaming affect the log($R$)--log($\lambda$) plots shown in Figure~\ref{fig:plot1}? The ratio of the core to extended radio luminosity, or flux density, is commonly used as an orientation indicator \citep[e.g.][]{orr82}. We would expect a beamed core to lift this ratio at small viewing angles to the jet axis. It is therefore possible that Figure~\ref {fig:plot3} suggests a net orientation difference between the upper and lower sequences in Figure~\ref{fig:plot1},  with sources on the lower track being viewed closer to pole-on on average. Indeed, SSL07 chose those SGs and LINERs for which at least the H$\alpha$ line is broad; in AGN orientation schemes \citep[e.g.][]{antonucci93}, broad lines are obscured by a torus of dense gas at large viewing angles to the jet. Hence, we might expect the lower track to be boosted upwards by a relatively larger amount, which would make any intrinsic bimodality less pronounced.

However, this interpretation is most likely far too simplistic. Firstly, though the presence of relativistic jets has been suggested for radio-quiet quasars \citep*[e.g.][]{falcke96,blundell03,barvainis05}, the situation appears to be less clear-cut for SGs, for which there is evidence for both sub-relativistic and relativistic expansion \citep[e.g.][]{ulvestad99,brunthaler00,middleberg04}. Moreover, recently \citet[][]{lal11} found no evidence for relativistic beaming in SGs. Furthermore, it has been proposed that SGs have thermally-dominated, baryonic jets with sub-relativistic velocities, though the jet velocity might be mildly relativistic close to the BH \citep[][]{bicknell98,bicknell02}. Therefore, at Eddington ratios $<1$ per cent, relativistic beaming may not be particularly significant for the lower sequence. 

We now attempt to quantify the possible beaming effects for the RGs and RLQs. In general, for a radio core with spectral index $\alpha$, the observed flux density $S_{\nu, \rm \: obs}$ is related to the intrinsic flux density $S_{\nu, \rm \: \rm rest}$ in the comoving frame by the expression \citep[][]{blandford79}       

\begin{equation}
S_{\nu, \rm \: obs}$ = $S_{\nu, \rm \: \rm rest} D^{k - \alpha},  
\end{equation}

where $k = 2$ for a continuous jet and $k=3$ for discrete ejections. The Doppler factor $D$ is determined using the Lorentz factor $\gamma$, dimensionless velocity $\beta = v/c$ and viewing angle to the jet $\theta$: 

\begin{equation}
D = [\gamma(1 - \beta \cos\theta)]^{-1}. 
\end{equation}
For the FR I RGs, bulk Lorentz factors are typically $\sim$2--10 \citep[][and references therein]{landt02}, and possibly higher for the BLRGs and RLQs \citep[e.g. $\gamma \sim$ $10$--$14$ in][]{mullin09}. Depending on the value of $\theta$ (in the range $0^{\circ}$--$90^{\circ}$ for the jet/counterjet), the core emission could be boosted or deboosted. For the FR I RGs, a fiducial viewing angle of $\sim$60$^{\circ}$ \citep[e.g.][and references therein]{landt02} could potentially lead to a decrease in radio loudness of up to several dex, depending on how relativistic the emitting electrons are. The gap between the sequences for log$(\lambda) < -2$ might therefore be intrinsically larger. On the other hand, the viewing angles of several sources in the FR I RGs sample have been estimated by \citet[][]{giovannini01}, and range from $<19^{\circ}$ to $\sim$85$^{\circ}$. Using the estimates of $\gamma$ and $\beta$ that are also available from their study, there are some cases where we would expect the core to be deboosted, others where the core should be boosted, and several where the parameters are not constrained sufficiently to allow us to differentiate between the two possibilities. At a given value of $\lambda$, a scatter of up to several dex in $R$ due to the effects of beaming is therefore certainly possible for the FR I RGs, as well as the BLRGs and RLQs. Indeed, the range in the unbinned RG radio loudness residuals for case (iii) in Figure~\ref{fig:plot2} is about 2.5 dex, for example. Beyond this, it is difficult to make any concrete predictions; with our current data we cannot rule out that the gap between the tracks is intrinsically larger, nor smaller. 

Another related issue is that there might be velocity structure in the jet, for example a fast spine surrounded by a slower sheath, with the core emission coming from the latter. As one might expect, the magnitude of the boosting/deboosting is smaller at lower values of $\gamma$. Such a scenario would reduce the importance of Doppler effects in the distribution of points in Figure~\ref{fig:plot1}.    

A further complication stems from evidence that the nuclear optical (and X-ray) emission in lower-power radio galaxies originates from a jet, instead of from the disc/accretion flow. For example, \citet*[][]{chiaberge99} found a strong correlation between the radio and optical nuclear flux densities for a sample of FR I RGs, suggesting a synchrotron origin for both \citep*[also see e.g.][]{chiaberge00,capetti00,hardcastle00,capetti02,hardcastle09}. Most of the FR I RGs in that paper were included in the SSL07 FR I RG subsample. The observed radio loudnesses will then depend additionally on the relative radio and optical Lorentz factors. However, more generally, the optical nuclear emission is potentially overestimated significantly (depending on the extent of the beaming for any jet-related emission). About one-third of the sources in the FR I RGs sample have upper limits only for $L_{B}$ (see Table~\ref{table: radio cores4}); this fraction may in fact be much larger. A decrease in $L_{B}$ would cause the points in the log($R$)--log($\lambda$) plane to move diagonally to the upper left along lines with a gradient of $-1$. Therefore, the gap in radio loudness between the upper and lower sequences would increase. On the other hand, using infrared data, SSL07 suggested that the accretion luminosities of some of the FR I RGs could in fact be underestimated by a factor of $\gtrsim$ 1 dex if the optical emission from the jet originates from outside an obscuring torus, with the core itself hidden by the obscuring material \citep[also see][]{cao04}. Correcting for this would result in the datapoints moving to the lower right along lines with a gradient of $-1$, closing rather than opening the gap between the sequences.

\subsection{Is BH spin alone responsible for the dichotomy?}

For log($\lambda$) $< -2$, the analysis in Section~\ref{core_vs_lobe} implies that Figures~\ref{fig:plot1} and \ref{fig:plot2} are essentially comparisons between the radio loudnesses of RGs (mainly FR I RGs) and SGs/LINERs (mainly SGs) that typically have a significant extended radio component which is associated with the active nucleus, though less so in the latter case, especially as there may also be contributions from starbursts. Moreover, the radio linear sizes of extended SGs are typically much smaller than for FR I RGs (e.g. $\sim$ a few kpc compared with $\sim$ hundreds of kpc). The fact that the FR I RGs are much more radio-loud than the SGs when total radio powers are used suggests a link between radio loudness bimodality and the presence of extended emission, which is similar to what was found by \citet{terashima03} and \citet{rafter11}, as well as the spatial scale of the extended emission. Yet if the `revised BH spin paradigm' presented by SSL07 is correct, then BH spin appears to affecting the extended emission substantially more than the core emission. Is this plausible?

The radio luminosity of extended emission scales with BH mass, external density and the age of the jet \citep[e.g.][]{heinz02}. Therefore, in panel (i) of Figure~\ref{fig:plot1}, the observed radio loudness dichotomy could at least in part be caused by the RGs residing in denser large-scale environments than the SGs and LINERs, in addition to the possible effects resulting from the interstellar medium (which is well known to be different for ellipticals and spirals) through which the jet propagates, as well as possible additional dependences on BH mass, jet lifetime, and, as advocated by SSL07, BH spin. Indeed, while low-redshift FR I RGs are well known to reside in rich environments on average \citep[e.g.][]{prestage88},\footnote{Also see \url{http://www.jb.man.ac.uk/atlas/} and references therein for the specific details of some of the sources in the SSL07 FR~I RG subsample.} this does not appear to be the case for SGs \citep*[e.g.][]{derobertis98}. Furthermore, one of the conclusions from the study of SG narrow-line region emission by \citet[][]{bicknell98} was that the ratio of the radio power to jet energy flux is much smaller than what is usually assumed for RGs, and that this is partially due to the smaller ages ($\sim$10$^{6}$ yr) of SGs compared with RGs. Also, it is interesting to note that in simulations of the growth and evolution of Fanaroff--Riley type II (FR II) radio galaxies by \citet[][]{wang08} and Kapi{\'n}ska et al. (in prep.), parameters such as jet age and environmental density are important, yet a jet kinetic luminosity that is strongly super-Eddington (i.e. possibly resulting from spin) is not required.

Whatever the underlying physical mechanisms that result in the bimodality in panel (i) of Figure~\ref{fig:plot1}, there is still an average inferred difference in jet power of about an order of magnitude at Eddington ratios $<1$ per cent for the mass-corrected, core radio loudnesses. Nonetheless, this difference is far less than the typical power variations of several orders of magnitude that are possible for realistic BH spin distributions, though the spread may also depend on the thickness of the accretion disc \citep[][]{meier01,tchekhovskoy10}: thinner discs result in a narrower range in power for a given spin distribution. Thus, depending on the accretion disc geometry, and provided that the mass-corrected core radio luminosity is indeed a reliable measure of jet power, then the dependence of jet power on BH spin may be much weaker than previously thought. This in turn reinforces our hypothesis that the ambient environment and/or jet lifetime could play a significant role in producing the dichotomy observed in total radio luminosity. Alternatively, our relatively small sample at low Eddington ratios may not be accurately representative of the general difference between the spins of BHs in giant ellipticals and spiral galaxies, which could be more extreme than is suggested here. In addition, as remarked by \citet[][]{sansigre11}, a dichotomy in BH spin may not necessarily lead to a clear dichotomy in radio loudness, given the range of possible uncertainties and selection effects. 

\begin{figure}
\epsfig{file=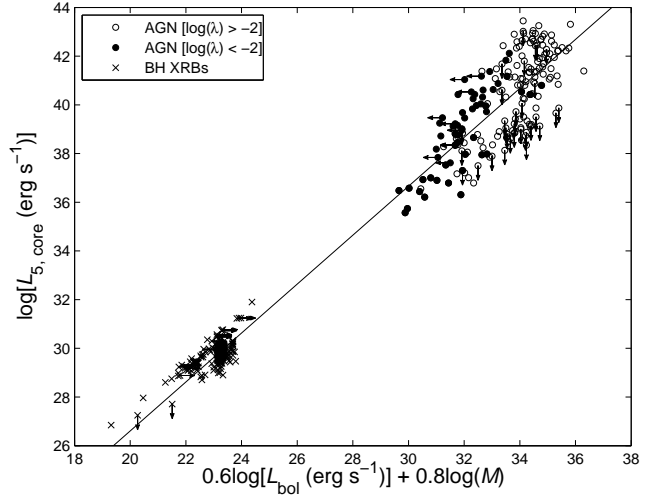,width=8.75cm}
\caption{Revised version of the fundamental plane of BH activity from \citet[][]{merloni03}, with X-ray luminosities replaced with radiative bolometric luminosities. The AGN data are from this paper and SSL07, while the BH XRB data are from \citet{fender10} and references therein, as well as \citet{calvelo10}. We assumed that log($M$) $=1$ for those XRBs whose BH masses have not yet been measured. The line of best fit is also shown.}
\label{fig:plot5}
\end{figure}

\subsection{Other potential uncertainties in our analysis}\label{potential uncertainties}

There are various sources of scatter that could help to blend the sequences in panels (ii) and (iii) in Figure~\ref{fig:plot1}. These include deviations about the mass correction term, the effects of relativistic beaming discussed previously, and uncertainties in the determinations of $L_{5, \rm \: core}$, $L_{B}$ and $M$. Moreover, additional scatter could be caused by the fact that the radio and optical nuclear measurements, which could be or are already known to be variable, were not conducted simultaneously. Some of the radio core luminosities may also be overestimated if the angular resolution was not sufficient to isolate the true radio core (itself a rather hard-to-define concept).

It is important to emphasise that although the fundamental planes of BH activity in \citet[][]{merloni03} and \citet[][]{falcke04} were defined using X-ray luminosities, we calculated radio loudnesses in this paper using $B$-band data. If the relationship between optical and X-ray core luminosity is complex, perhaps due to dust extinction, for example, then further scatter will be introduced to the plots in Figure~\ref{fig:plot1}; \citet[][]{terashima03} discuss some other issues that may affect radio loudnesses derived using optical data. On the other hand, Figure~\ref{fig:plot5} suggests that
our assumption in Section~\ref{AGN samples} of a simple relationship between $L_{X}$ and $L_{B}$ is reasonable in general. In this figure, we have plotted a revised version of the fundamental plane of BH activity shown in \citet[][]{merloni03}, replacing the X-ray luminosity term on the horizontal axis with a corresponding term for the radiative bolometric luminosity. The data points include the AGN studied in this paper, as well as a list of core measurements for BH XRBs in the low/hard state \citep*[][]{fender10,calvelo10}. For the AGN, the bolometric correction is assumed to be $L_{\rm bol} = 10 L_{B}$ (see Section~\ref{introduction}), while for the BH XRBs we used the bolometric correction for the low/hard state from \citet[][]{migliari06}: $L_{\rm bol} = 5 L_{X}$. The distribution of points in Figure~\ref{fig:plot5} agrees well overall with the distribution in the fundamental plane plot from \citet[][]{merloni03}. The amount of scatter observed for the AGN depends on whether we exclude those sources with log$(\lambda) > -2$, where there may be a dependence on accretion state (see Section~\ref{introduction}). The scatter amongst the XRBs is discussed in both \citet[][]{fender10} and \citet[][]{calvelo10}. The slope of the line of best fit to all data points (treating limits as detections as before) is 1.01; the slope varies by at most a few percent if we remove AGN with log($\lambda$) $> -2$, or treat the limits more rigorously. Therefore, the correlation is close to the form $L_{5, \rm \: core} \propto L_{\rm bol}^{0.6} M^{0.8}$, which is very similar to equation~\ref{eqn_merloni}.

\begin{figure}
\epsfig{file=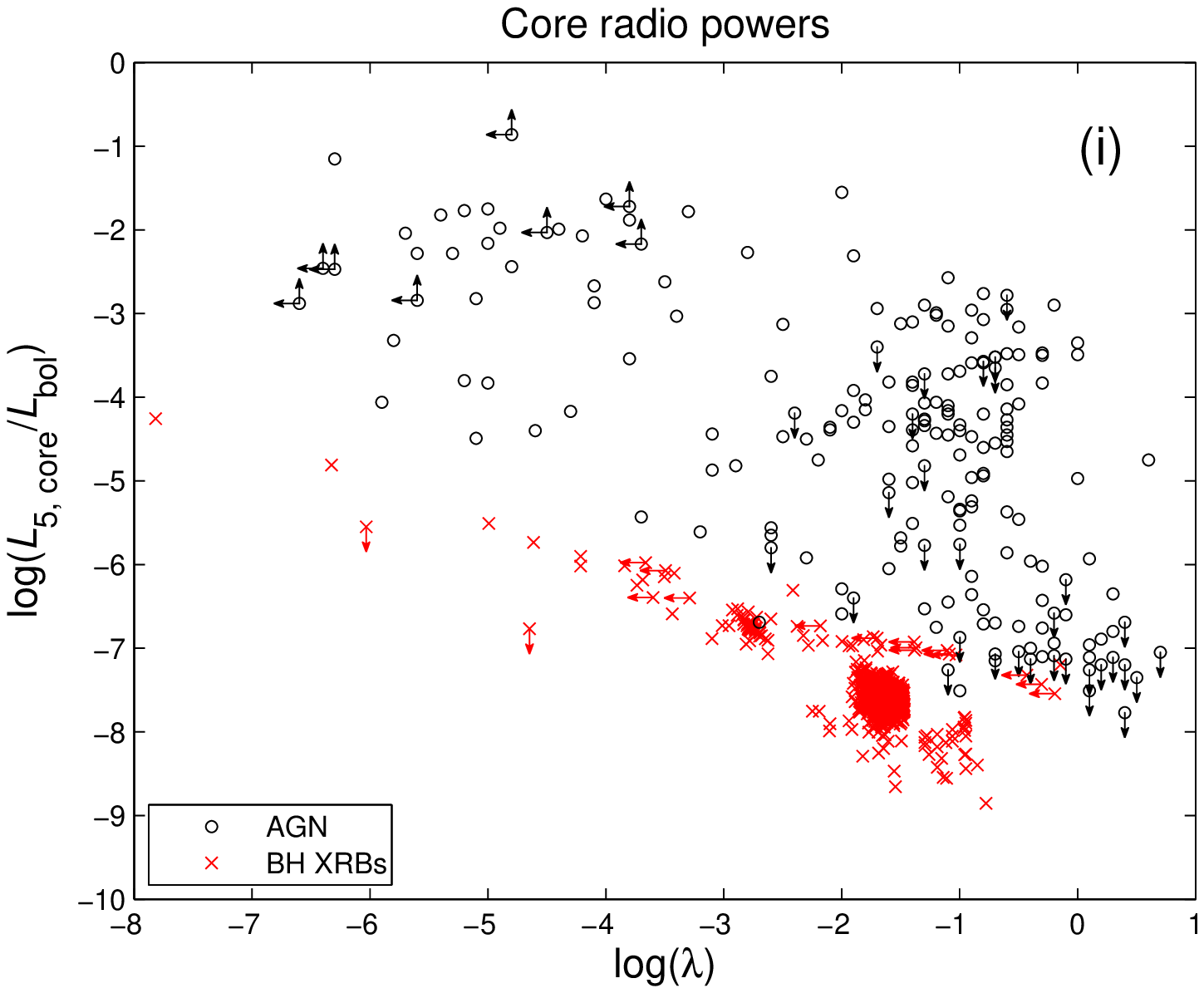,width=8.40cm}
\epsfig{file=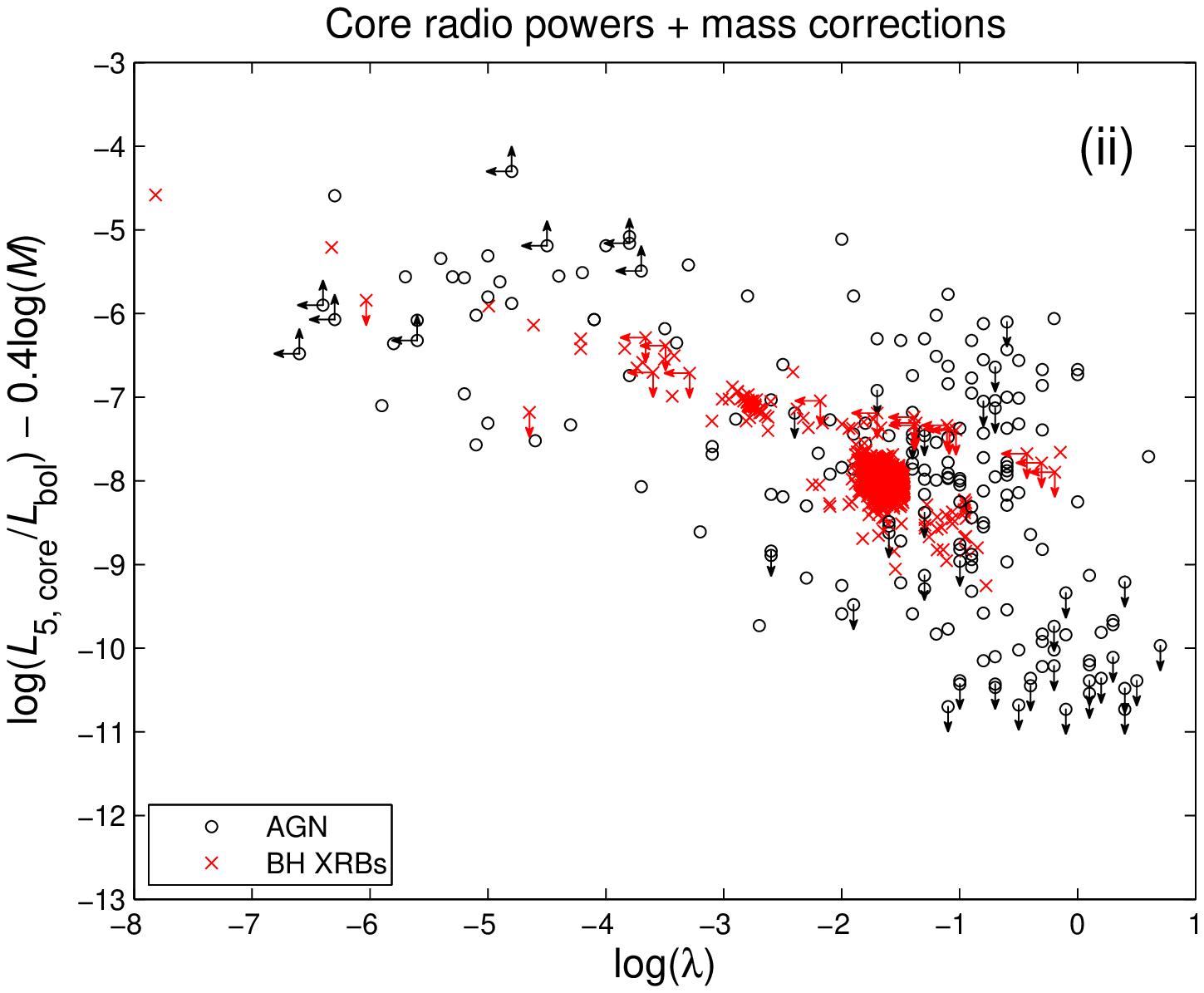,width=8.40cm}
\caption{Logarithm of the ratio of core radio to total radiative bolometric luminosity, plotted as a function of log($\lambda$). In panel (ii), the mass correction term has been applied on the vertical axis. The BH XRB data are the same used in Figure~\ref{fig:plot5}; the cluster of XRB points seen in both panels contains the measurements for Cygnus X--1.}
\label{fig:plot4}
\end{figure}

\subsection{The importance of the mass correction term}

Figure~\ref{fig:plot5} shows how a mass term allows the BH XRB and AGN populations to be linked. From Figure~\ref{fig:plot1}, it is also clear that the 0.4log($M$) mass term correction plays an important role in collapsing the radio loudness dichotomy. This is further explored in Figure~\ref{fig:plot4}, where we have also included the BH XRBs, which have much lower BH masses than the AGN, in the log($R$)--log($\lambda$) plots. To eliminate the need to extrapolate from X-rays to $B$-band for the BH XRBs, and vice-versa for the AGN, we have used a different measure of radio loudness in Figure~\ref{fig:plot4}, namely the ratio of the core radio luminosity to total radiative bolometric luminosity. Note that Section~\ref{potential uncertainties} not only implies that our bolometric corrections appear to be sensible, but that the mass term is approximately 0.4log($M$).

In panel (i) of Figure~\ref{fig:plot4}, the BH XRBs appear to trace out a third track (and possibly a fourth; see \citealt{fender10} and \citealt[][]{calvelo10}) below the two AGN sequences. However, once the mass correction is applied, the BH XRBs are shifted towards the middle of the AGN distribution. This powerfully illustrates that the spread in radio loudness is reduced significantly once the dependence on BH mass is taken into account.
\newline
\newline

\subsection{What can we learn from BH XRBs?}

Given that a link is implied by the distribution in panel (ii) of Figure~\ref{fig:plot4}, studies of BH XRBs can be used shed further light on how BH spin may affect AGN jet power. There is in fact no compelling evidence linking the observed properties of the jets in BH XRBs with the reported measurements of BH spin \citep[][]{fender10}. Furthermore, as remarked by \citet[][]{fender10}, the apparently low-spin BH in Cygnus X-1 is likely to be associated with strong extended radio emission. These results cast further doubt on the importance of spin in the powering of AGN jets (but see discussion in \citealt{fender10} and \citealt{sansigre11}). Instead of causing of a dichotomy in radio loudness, spin may just be an additional source of scatter in the log($R$)--log($\lambda$) plane, or perhaps it has no effect at all.

\section{Conclusions}\label{conclusions}

In this paper, we have used a previously published sample of 199 sources to gain new insights into the radio-loud/radio-quiet dichotomy that is claimed to exist for AGN. Our main findings are as follows:

\begin{enumerate}

\item When radio loudness is plotted as a function of Eddington ratio, the distribution of data points is strongly dependent on the measure of radio power adopted. The clear bimodality that is present when total radio powers are used becomes much less apparent when the radio loudness is determined from the core only, and a mass correction is applied.
\item For mass-corrected, core radio loudnesses and Eddington ratios $< 1$ per cent, FR I RGs and BLRGs are on average more radio-loud than SGs and LINERs by about 1.6 dex, or equivalently have total jet powers about 1 dex stronger. This offset is much smaller than what could be expected if BH spin was the primary mechanism powering AGN jets.
\item If BH spin is in fact not as important as previously thought, the radio loudness dichotomy seen in total radio luminosity could well depend on several factors in addition to, or instead of spin: the environmental density, the age of the radio source, and the dependence of radio emission strength on black hole mass.  
\end{enumerate}

So, in summary, a mild radio loudness dichotomy for AGN appears to persist despite the implementation of core radio luminosities and a mass term correction, although these adjustments significantly reduce the required range of jet powers. Simultaneous radio and optical (or X-ray) nuclear measurements, the former done with Very Long Baseline Interferometry (VLBI), for a large, carefully-selected sample would help considerably to confirm if there truly is a dichotomy in AGN jet power.

\section*{Acknowledgments}

We thank the referee for several helpful suggestions which improved this paper. We also thank Phil Uttley, Elmar K{\"o}rding and Andrea Merloni for stimulating discussions. JWB acknowledges support from The Science and Technology Facilities Council in the United Kingdom.  This research has made use of the NASA/IPAC Extragalactic Database (NED) which is operated by the Jet Propulsion Laboratory, California Institute of Technology, under contract with the National Aeronautics and Space Administration. In addition, this research has made use of the VizieR catalogue access tool (CDS, Strasbourg, France), and Ned Wright's Javascript Cosmology Calculator \citep{wright06}.
\newline
\newline

{}

\appendix

\section{Core radio luminosities and radio loudness parameters for the SSL07 AGN sample}\label{appendix radio fluxes}

For the five subsamples from SSL07, core radio luminosities (in the form $\nu_{5} L_{\nu_{5, \rm \: core}}$) and the resulting radio loudness parameters ($L_{\nu_{5, \rm \: core}}/L_{\nu_{B}}$) are presented in Tables~\ref{table: radio cores1} to \ref{table: radio cores5}.

\begin{table*}
\caption{Core radio luminosities and radio loudness parameters for the subsample of broad-line radio galaxies.}
\setlength{\tabcolsep}{3pt}
\label{table: radio cores1}
\begin{center}
\begin{tabular}{lrrcclrrc}
\hline
\multicolumn{1}{c}{Source name} & \multicolumn{1}{c}{log($L_{5, \rm \: core}$)} & \multicolumn{1}{c}{log($R$)} & References &  &  \multicolumn{1}{c}{Source name} & \multicolumn{1}{c}{log($L_{5, \rm \: core}$)} & \multicolumn{1}{c}{log($R$)} & References \\
& \multicolumn{1}{c}{(erg s$^{-1}$)} & & & & & \multicolumn{1}{c}{(erg s$^{-1}$)} &  &  \\
\hline
3C 17	& $	42.6	$ & $	3.82	$ & 	MKT93	&	&		PKS 1254$-$333	& $	40.3	$ & $	1.56	$ & 	RKP99	\\	
4C $+$11.06	& $	41.1	$ & $	2.41	$ & 	LGH94	&		&	3C 287.1	& $	42.4$\rlap{$^f$}	 & $	3.57	$ & 	Do86	 \\	
3C 59	& $	40.4	$ & $	1.16	$ & 	LM97	&	&		Mrk 0668	& $	<39.4$\rlap{$^g$}	 & $	<0.37	$ & 	St97	\\	 
3C 67	& $	<40.6	$ & $	<1.93	$ & 	Sa95, Gil04	&		&	PKS 1417$-$19	& $	38.8	$ & $	-0.32	$ & 	ZB95	\\	
IRAS 02366$-$3101	& $	39.4$\rlap{$^a$}	 & $	0.35	$ & 	Ro94	&		&	3C 303	& $	41.5	$ & $	3.01	$ & 	BP84, Gio88, LM97	\\	
PKS 0236$+$02	& $	41.1	$ & $	2.11	$ & 	LM97	&		&	PKS 1514$+$00	& $	41.4	$ & $	3.12	$ & 	LM97	\\	 
B2 0309$+$39	& $	41.1	$ & $	2.27	$ & 	WB86, LM97		&	&	4C $+$35.37	& $	<39.3\rlap{$^g$}	$ & $	<0.37	$ & 	 An85	\\	
PKS 0340$-$37	& $	<41.8	$ & $	<2.73	$ & 	BH06	&		&	3C 332	& $	40.4	$ & $	1.66	$ & 	Gio88	\\	
3C 93	& $	40.8	$ & $	1.64	$ & 	Ka95	&	&		MC2 1635$+$119	& $	40.8	$ & $	1.44	$ & 	LGH94, LM97	\\	
MS 0450.3$-$1817	& $	<39.0$\rlap{$^b$}	 & $	<1.94	$ & 	FMZ82	&		&	Arp 102B	& $	39.9	$ & $	2.38	$ & 	 LM97	\\	
Pictor A	& $	41.2$\rlap{$^c$}	 & $	3.01	$ & 	MKT93, JMR94	&		&	PKS 1739$+$184	& $	41.1	$ & $	1.97	$ & 	 LGH94 	\\	
CBS 74	& $	<40.1$\rlap{$^d$}	 & $	<0.99	$ & 	Be95	&	&		3C 382	& $	40.8	$ & $	1.83	$ & 	WB86, Gio88, LM97	\\	 
PKS 0846$+$101	& $	41.0	$ & $	1.79	$ & 	LGH94	&		&	3C 390.3	& $	41.1	$ & $	2.21	$ & 	WB86, Gio88 	\\	 
PKS 0857$-$191	& $	42.4\rlap{$^e$}	$ & $	3.24	$ & 	Mu10	&		&	PKS 2058$-$425	& $	42.6\rlap{$^e$}	$ & $	3.18	$ & 	 Mu10, ATCA, VLA 	\\	
4C $+$05.38	& $	41.8	$ & $	2.32	$ & 	LM97	&		&	3C 445	& $	40.5	$ & $	1.87	$ & 	BP84, WB86, MKT93 	\\	 
PKS 0921$-$213	& $	40.0	$ & $	1.98	$ & 	ZB95	&		&	PKS 2300$-$18	& $	41.9	$ & $	3.20	$ & 	BP84, RKP99	 \\	
3C 227	& $	40.5	$ & $	1.78	$ & 	WB86, MKT93	&		&	PKS 2305$+$188	& $	42.0	$ & $	2.54	$ & 	LGH94	\\	
B2 1028$+$31	& $	41.5	$ & $	2.07	$ & 	LGH94, LM97		&	&	MC3 2328$+$167	& $	<41.7$\rlap{$^h$}	 & $	<2.62	$ & 	 LGH94	\\	
PKS 1151$-$34	& $	43.4$\rlap{$^e$} & $	4.58	$ & 	Pe82, RKP99, BH06		&	&		&		&		&		\\
	&  & 	 & 	ATCA, VLA		&	&		&		&		&		\\	
\hline
\multicolumn{9}{p{160mm}}{{\it References}: An85 -- \citet{antonucci85}; ATCA --  Australia Telescope Compact Array calibrator database (\url{http://www.narrabri.atnf.csiro.au/calibrators/}); Be95 -- \citet{becker95}; BH06 -- \citet{burgess06}; BP84 -- \citet{bridle84}; Do86 -- \citet{downes86}; FMZ82 -- \citet*{feigelson82}; Gil04 -- \citet{gilbert04}; Gio88 -- \citet{giovannini88}; JMR94 -- \citet*{jones94}; Ka95 -- \citet{kapahi95}; LGH94 -- \citet*{lister94}; LM97 -- \citet{lm97}; MKT93 -- \citet*{morganti93}; Mu10 -- \citet{murphy10}; Pe82 -- \citet{perley82}; RKP99 -- \citet*{reid99}; Ro94 -- \citet{roy94}; Sa95 -- \citet{sanghera95}; St97 -- \citet{stanghellini97}; VLA -- Very Large Array calibrator database (\url{http://www.aoc.nrao.edu/~gtaylor/csource.html)}; WB86 -- \citet{wills86}; ZB95 -- \citet{zirbel95}.}\\
\multicolumn{9}{p{160mm}}{{\it General note}: ATCA and VLA calibrator data were obtained from a range of array configurations.} \\
\multicolumn{9}{p{160mm}}{$^a$We extrapolated the 2.29 GHz core flux density to 5 GHz assuming that $\alpha=0$.} \\
\multicolumn{9}{p{160mm}}{$^b$Upper limit calculated using the integrated flux density of the total radio source in FMZ82.} \\
\multicolumn{9}{p{160mm}}{$^c$The core flux densities presented in JMR94 are at 2.3 and 8.4 GHz. To calculate the expected 5 GHz core flux density, we interpolated between these measurements, and the result was subsequently averaged with the core flux density from MKT93.} \\
\multicolumn{9}{p{160mm}}{$^d$5 GHz upper limit for the core estimated using the 1.4 GHz integrated flux density of the total radio source, and assuming that $\alpha=0$.} \\
\multicolumn{9}{p{160mm}}{$^e$Estimated fully or partly from the high-resolution integrated flux density of the total radio source, where no source extension was reported (except for the Pe82 measurement for PKS 1151$-$34 where slight source extension was noted).}\\
\multicolumn{9}{p{160mm}}{$^f$The core flux density was estimated from the peak flux density of the central component in the 5 GHz contour map.} \\
\multicolumn{9}{p{160mm}}{$^g$Core not detected; a $5\sigma$ upper limit was used.} \\
\multicolumn{9}{p{160mm}}{$^h$LGH94 remarked that the core flux density is probably contaminated by nearby extended structure. We have therefore used their measurement as an upper limit.} \\
\end{tabular}
\end{center}
\end{table*}

\begin{table*}
\caption{Core radio luminosities and radio loudness parameters for the subsample of radio-loud quasars.}
\label{table: radio cores2}
\setlength{\tabcolsep}{3.75pt}
\begin{center}
\begin{tabular}{lrrcclrrc}
\hline
\multicolumn{1}{c}{Source name} & \multicolumn{1}{c}{log($L_{5, \rm \: core}$)} & \multicolumn{1}{c}{log($R$)} & References &  &  Source name & \multicolumn{1}{c}{log($L_{5, \rm \: core}$)} & \multicolumn{1}{c}{log($R$)} & References \\
& \multicolumn{1}{c}{(erg s$^{-1}$)} & &  & & & \multicolumn{1}{c}{(erg s$^{-1}$)} &  &  \\
\hline
4C 25.01	& $	42.5	$ & $	2.00	$ & 	LGH94	&	&	B2 1223$+$25	& $	40.5	$ & $	0.80	$ & 	LGH94	\\
B2 0110$+$29	& $	42.1	$ & $	2.04	$ & 	LM97	&	&	PKS 1233$-$24	& $	41.3	$ & $	1.19	$ & 	RKP99	\\
4C 31.06	& $	42.7	$ & $	3.04	$ & 	LM97	&	&	3C 277.1	& $	41.6	$ & $	2.06	$ & 	WB86	\\
PKS 0202$-$76	& $	42.6	$ & $	2.44	$ & 	BH06	&	&	PKS 1302$-$102	& 	$42.9$	 &  $	2.31	$ & 	 Ke89, MRS93, LGH94,	\\
PKS 0214$+$10	& $	42.1	$ & $	1.86	$ & 	Ka95, LM97            &       &                       &                 &               &       ATCA, VLA       \\
PKS 0312$-$77	& $	42.7\rlap{$^a$}	$ & $	2.64	$ & 	Mu10	&	&	PKS 1346$-$112	& $	42.9$\rlap{$^{ac}$}	 & $	3.24	$ & 	 WPW03, He07	\\
3C 111	& $	41.5	$ & $	1.53	$ & 	BP84, LM97	&	&	B2 1351$+$26	& $	41.6	$ & $	1.79	$ & 	WB86, LM97	\\
PKS 0558$-$504	& $	41.4\rlap{$^a$}	$ & $	1.39	$ & 	Dr03, Mu10	&	&	PKS 1355$-$41	& $	42.2$\rlap{$^d$}	 & $	1.85	 $ & 	MKT93, JMR94, BH06	\\
B2 0742$+$31	& $	43.3	$ & $	2.55	$ & 	BP84, LGH94	&	&	CSO 0643	& $	42.1	$ & $	2.31	$ & 	LM97	 \\
PKS 0812$+$02	& $	42.7	$ & $	2.61	$ & 	BP84, LM97	&	&	PKS 1451$-$375	& $	43.3	$ & $	3.19	$ & 	 WB86, ATCA, VLA	\\
3C 206	& $	41.9	$ & $	1.93	$ & 	Ka95, RKP99	&	&	4C $+$37.43	& $	42.2	$ & $	1.93	$ & 	WB86, Ke89, MRS93	\\
PKS 0925$-$203	& $	43.1\rlap{$^a$}	$ & $	2.79	$ & 	Pe82, ATCA, VLA	&	&	LB 9743	& $	41.5	$ & $	1.60	$ & 	LGH94	 \\
4C $+$09.35	& $	40.9	$ & $	1.18	$ & 	LGH94	&	&	4C $+$18.47	& $	41.3	$ & $	1.68	$ & 	Ka95, LM97	\\
PKS 1004$-$217	& $	42.7\rlap{$^a$}	$ & $	2.64	$ & 	Mu10	&	&	3C 351	& $	41.4	$ & $	0.62	$ & 	WB86, Ke89, MRS93	 \\
PKS 1004$+$13	& $	41.2	$ & $	0.82	$ & 	BP84, Ke89, MRS93	&	&	B2 1719$+$35	& $	42.6	$ & $	2.98	$ & 	 LM97	 \\
                &                 &               &                 LGH94, Ka95                     &       & B2 1721$+$34	& $	42.3	 $ & $	2.66	$ & 	LGH94	\\
PKS 1011$-$282	& $	41.4	$ & $	1.59	$ & 	RKP99	&	&	PKS 1725$+$044	& $	42.9	$ & $	2.97	$ & LGH94, LM97	\\
PKS 1020$-$103	& $	<42.3\rlap{$^{b}$}	$ & $	<2.42	$ & 	LGH94	&	&	MRC 1745$+$163	& $	<42.5$\rlap{$^e$}	 & $	 <2.49	$ & 	GC91	\\
3C 246	& $	41.9	$ & $	1.81	$ & 	Ke89, MRS93, Ka95	&	&	PKS 1914$-$45	& $	42.2\rlap{$^a$}	$ & $	2.06	$ & 	Mu10	 \\
PKS 1101$-$325	& $	42.9	$ & $	2.64	$ & 	ATCA, VLA	&	&	PKS 2140$-$048	& $	42.9	$ & $	3.37	$ & 	 ATCA, VLA	\\
B2 1104$+$36	& $	41.4	$ & $	1.70	$ & 	LM97	&	&	OX $+$169	& $	42.5	$ & $	2.29	$ & 	LGH94, LM97	\\
B2 1128$+$31	& $	41.9	$ & $	1.77	$ & 	LM97	&	&	PKS 2208$-$137	& $	42.8	$ & $	2.66	$ & 	WB86, RKP99	\\
PKS 1146$-$037	& $	42.6$\rlap{$^{ac}$}	 & $	2.85	$ & WPW03, He07	&	&	PKS 2227$-$399	& $	42.7\rlap{$^a$}	$ & $	3.14	 $ & 	Pe82, ATCA, VLA	\\
LB 2136	& $	42.9	$ & $	3.06	$ & 	BP84, LM97	&	&	PKS 2247$+$14	& $	<43.0$\rlap{$^b$}	 & $	<3.35	$ & 	LGH94	\\
TXS 1156$+$213	& $	<40.9	$ & $	<1.31	$ & 	WB86	&	&	PKS 2302$-$713	& $	<42.4$\rlap{$^f$}	 & $	<2.57	$ & 	 JMR94	 \\
B2 1208$+$32A	& $	41.1	$ & $	0.89	$ & 	LM97	&	&	PKS 2349$-$01	& $	41.7	$ & $	1.74	$ & 	WB86	\\
\hline
\multicolumn{9}{p{160mm}}{{\it References}: ATCA --  Australia Telescope Compact Array calibrator database (\url{http://www.narrabri.atnf.csiro.au/calibrators/}); BH06 -- \citet{burgess06}; BP84 -- \citet{bridle84}; Dr03 -- \citet{drake03}; GC91 -- \citet{gregory91}; He07 -- \citet{healey07}; JMR94 -- \citet{jones94}; Ka95 -- \citet{kapahi95}; Ke89 -- \citet{kellermann89}; LGH94 -- \citet{lister94}; LM97 -- \citet{lm97}; MKT93 -- \citet{morganti93}; MRS93 -- \citet{miller93}; Mu10 -- \citet{murphy10}; Pe82 -- \citet{perley82}; RKP99 -- \citet{reid99}; VLA -- Very Large Array calibrator database (\url{http://www.aoc.nrao.edu/~gtaylor/csource.html)}; WB86 -- \citet{wills86}; WPW03 -- \citet*{winn03}.}  \\
\multicolumn{9}{p{160mm}}{{\it General notes}: ATCA and VLA calibrator data were obtained from a range of array configurations. Also, if there were more than two measurements for a particular source, and two of these were from Ke89 and MRS93, then the flux densities from these two studies were averaged together first and henceforth treated as a single measurement. This is because MRS93 reanalysed the data presented in Ke89.} \\
\multicolumn{9}{p{160mm}}{$^a$Estimated fully or partly from the high-resolution integrated flux density of the total radio source, where no source extension was reported.} \\
\multicolumn{9}{p{160mm}}{$^b$LGH94 remarked that the core flux density is probably contaminated by nearby extended structure. We therefore used their measurement as an upper limit.} \\
\multicolumn{9}{p{160mm}}{$^c$The flux densities presented in WPW03 and He07 are at 8.4 GHz; we have estimated the 5 GHz core flux density by assuming that $\alpha=0$.} \\
\multicolumn{9}{p{160mm}}{$^d$The flux densities presented in JMR94 are at 2.3 and 8.4 GHz; to calculate the expected 5 GHz core flux density, we interpolated between these measurements.} \\
\multicolumn{9}{p{160mm}}{$^e$Upper limit calculated using the integrated flux density of the total radio source in GC91.} \\
\multicolumn{9}{p{160mm}}{$^f$We extrapolated the 2.3 GHz upper limit for the core flux density to 5 GHz assuming that $\alpha=0$.} \\

\end{tabular}
\end{center}
\end{table*}

\begin{table*}
\caption{Core radio luminosities and radio loudness parameters for the subsample of Seyfert and low-ionization nuclear emission-line region galaxies.}
\label{table: radio cores3}
\begin{center}
\begin{tabular}{lrrcclrrc}
\hline
\multicolumn{1}{c}{Source name} & \multicolumn{1}{c}{log($L_{5, \rm \: core}$)} & \multicolumn{1}{c}{log($R$)} & References &  &  Source name & \multicolumn{1}{c}{log($L_{5, \rm \: core}$)} & \multicolumn{1}{c}{log($R$)} & References \\
& \multicolumn{1}{c}{(erg s$^{-1}$)} & &  & & & \multicolumn{1}{c}{(erg s$^{-1}$)} &  &  \\
\hline
Mrk 335	& $	38.4	$ & $	0.60	$ & Ke89, MRS93		&	&		NGC 4258	& $	35.7	$ & $	2.07	$ &	HU01, NFW05	\\
Fairall 9	& $	<39.7\rlap{$^a$}	$ & $	<-0.05	$ &	Ho02	&		&	NGC 4388	& $	36.9	$ & $	1.26	$ &	HU01	\\
Mrk 590	& $	38.4\rlap{$^b$}	$ & $	1.38	$ &	UW84a	&	&		NGC 4565	& $	36.2\rlap{$^b$}		$ & $	1.64	$ &	HU01, NFW05	\\
NGC 1068	& $	38.8	$ & $	1.74	$ &	HU01	&		&	NGC 4579	& $	37.5	$ & $	1.96	$ &	AUH04, NFW05	\\
Ark 120	& $	38.6\rlap{$^c$}	$ & $	-1.02	$ &	BA89	&		&	NGC 4593	& $	37.2\rlap{$^b$}	$ & $	-0.40	$ &	UW84b	\\
Mrk 79	& $	37.8\rlap{$^c$}	$ & $	-0.62	$ &	UW84a	&	&		NGC 4639	& $	35.6\rlap{$^b$}	$ & $	0.70	$ &	HU01	\\
NGC 2787	& $	36.6	$ & $	2.81	$ &	NFW05	&	&		NGC 5033	& $	36.8	$ & $	0.52	$ &	HU01	\\
Mrk 110	& $	38.4	$ & $	0.17	$ &	Ke89, MRS93	&		&	NGC 5252	& $	38.8	$ & $	2.59	$ &	WT94	\\
NGC 3031	& $	36.9	$ & $	1.73	$ &	HU01	&	&		NGC 5273	& $	36.4\rlap{$^b$}	$ & $	0.57	$ &	HU01	\\
NGC 3227	& $	36.3		$ & $	-0.56	$ &	NFW05	&		&	IC 4329A	& $	38.8	$ & $	0.67 $ &	Un87	\\
NGC 3516	& $	37.2	$ & $	-0.16	$ &	NFW05	&		&	Mrk 279	& $	38.4\rlap{$^d$}	$ & $	-0.41	$ &	Ku95	\\
Mrk 744	& $	37.0\rlap{$^d$}	$ & $	-0.46	$ &	Ku95	&		&	NGC 5548	& $	37.9	$ & $	0.48	$ &	HU01	\\
NGC 3783	& $	38.1\rlap{$^b$}	$ & $	-0.01	$ &	UW84b	&	&		Mrk 817	& $	38.5\rlap{$^c$}	$ & $	0.45	$ &	BA89	\\
NGC 3982	& $	36.5\rlap{$^b$}		$ & $	1.31	$ &	HU01	&		&	Mrk 841	& $	38.0	$ & $	0.21	$ &	Ke89, MRS93	\\
NGC 3998	& $	38.0	$ & $	2.30	$ &	NFW05	&		&	NGC 5940	& $	<37.5\rlap{$^d$}	$ & $	<-0.27	$ &	 Ku95	 \\
NGC 4051	& $	36.5	$ & $	0.08	$ &	HU01	&		&	NGC 6104	& $	<37.3\rlap{$^d$}	$ & $	<0.33	$ &	Ku95	\\
NGC 4151	& $	36.8\rlap{$^e$}	$ & $	-1.38	$ &	NFW05	&		&	Mrk 509	& $	38.5\rlap{$^c$}	$ & $	-0.97	$ &	SW92	\\
NGC 4203	& $	37.0		$ & $	2.33	$ &	AUH04, NFW05	&		&	NGC 7469	& $	38.6\rlap{$^d$}	$ & $	0.94	$ &	Ku95	\\
Mrk 766	& $	38.3\rlap{$^c$}	$ & $	1.11	$ &	UW84a	&		&	Mrk 530	& $	38.7\rlap{$^f$}	$ & $	1.69	$ &	Ro94, Ku95	 \\
\hline
\multicolumn{9}{p{160mm}}{{\it References}: AUH04 -- \citet*[][]{anderson04}; BA89 -- \citet[][]{barvainis89}; Ho02 -- \citet{ho02}; HU01 -- \citet{houl01}; Ke89 -- \citet{kellermann89}; Ku95 -- \citet{kukula95}; MRS93 -- \citet{miller93}; NFW05 -- \citet*[][]{nagar05}; Ro94 -- \citet{roy94}; SW92 -- \citet[][]{singh92}; Un87 -- \citet[][]{unger87}; UW84a -- \citet{ulvestad84a}; UW84b -- \citet{ulvestad84b}; WT94 -- \citet[][]{wilson94}.}  \\
\multicolumn{9}{p{160mm}}{{\it General note}: HU01 and NFW05 presented both peak and integrated core flux densities for the sources in their samples. If HU01 or NFW05 is listed as the reference for a particular source, we note that we used the peak core flux density (or the peak total flux density where necessary). } \\
\multicolumn{9}{p{160mm}}{$^a$Ho02 reported an upper limit for the `nuclear' radio luminosity; we have used this measurement as an upper limit for the core luminosity as well.} \\
\multicolumn{9}{p{160mm}}{$^b$Estimated from the high-resolution integrated or peak flux density of the total radio source (except in NFW05 for NGC 4565), where no source extension was reported.} \\
\multicolumn{9}{p{160mm}}{$^c$Estimated from the peak flux density of the total radio source, which is either partially extended or more fully extended. If more than one 6 cm contour map was available for a particular source, we used the highest-resolution version.} \\
\multicolumn{9}{p{160mm}}{$^d$Estimated from the peak flux density of the total radio source at 8.4 GHz, or the quoted upper limit for the 8.4 GHz flux density; we extrapolated this value to 5 GHz assuming that $\alpha=0$.}\\
\multicolumn{9}{p{160mm}}{$^e$NFW05 estimated the 5 GHz core flux density using 1.4 GHz data.}\\
\multicolumn{9}{p{160mm}}{$^f$Estimated from interpolating between the core flux density at 2.3 GHz (Ro94) and the peak flux density of the total radio source at 8.4 GHz (Ku95).}\\
\end{tabular}
\end{center}
\end{table*}

\begin{table*}
\caption{Core radio luminosities and radio loudness parameters for the subsample of FR I radio galaxies.}
\setlength{\tabcolsep}{3.5pt}
\label{table: radio cores4}
\begin{center}
\begin{tabular}{lrrcclrrc}
\hline
\multicolumn{1}{c}{Source name} & \multicolumn{1}{c}{log($L_{5, \rm \: core}$)} & \multicolumn{1}{c}{log($R$)} & References & &  Source name & \multicolumn{1}{c}{log($L_{5, \rm \: core}$)} & \multicolumn{1}{c}{log($R$)} & References \\
& \multicolumn{1}{c}{(erg s$^{-1}$)} & &  & & & \multicolumn{1}{c}{(erg s$^{-1}$)} &  &  \\
\hline
3C 28	&	$<39.0$	&	$\cdots$\rlap{$^a$}	&	Gio88	&	&		NGC 4874	&	$37.8$	&	$>3.68$\rlap{$^b$}		&	Gio88	\\
3C 29	&	$40.3$	&	$4.16$	&	MKT93	&	&		3C 288	&	$41.4$	&	$4.51$		&	Gio88, LM97	\\
NGC 315	&	$40.2$	&	$4.38$	&	BP84, Gio88, LM97	&		&	3C 296	&	$39.7$	&	$4.32$		&	BP84, Gio88	\\
3C 31	&	$39.5$	&	$3.70$	&	BP84, Gio88	&		&	3C 310	&	$40.4$	&	$4.25$		&	Gio88 	\\
NGC 507	&	$37.6$	&	$>3.25$\rlap{$^b$}	&	Gio88	&	&		3C 314.1	&	$<39.2$	&	$\cdots$\rlap{$^a$}		&	Gio88	\\
3C 40	&	$39.5$	&	$>4.11$\rlap{$^b$}	&	BP84, MKT93	&		&	3C 317	&	$40.6$	&	$4.36$		&	MKT93, LM97	\\
NGC 741	&	$38.4$	&	$>3.29$\rlap{$^b$}	&	LM97	&	&		3C 338	&	$40.0$	&	$3.97$	&	Gio88, LM97	\\
3C 66B	&	$39.8$	&	$3.46$	&	BP84, Gio88 	&		&	3C 346	&	$41.8$	&	$3.87$		&	Gio88	\\
3C 78	&	$40.9$	&	$3.51$	&	BP84, MKT93	&		&	3C 348	&	$40.6$	&	$4.15$		&	BP84, MKT93, GL03	\\
3C 83.1	&	$39.2$	&	$4.10$	&	BP84, Ch99	&		&	3C 424	&	$40.5$\rlap{$^c$}	&	$>3.96$\rlap{$^b$}		&	Ch99	\\
3C 84	&	$42.1$	&	$4.35$	&	NdB82, BP84, LM97, Ch99		&	&	3C 438	&	$41.2$	&	$>4.41$\rlap{$^b$}		&	BP84, Ch99	 \\
3C 89	&	$41.0$	&	$>5.28$\rlap{$^b$}	&	Ch99	&	&		3C 442	&	$38.2$	&	$3.32$		&	Gio88	\\
3C 264	&	$40.0$	&	$3.11$	&	Gio88	&	&		3C 449	&	$39.1$	&	$3.27$		&	BP84, Gio88	\\
3C 270	&	$39.3$	&	$4.99$	&	MKT93, LM97	&		&	NGC 7626	&	$38.3$	&	$>3.67$\rlap{$^b$}		&	BP84	\\
3C 272.1	&	$38.7$	&	$3.85$	&	BP84, Gio88		&	&	3C 465	&	$40.4$	&	$4.06$		&	BP84, Gio88	\\
3C 274	&	$39.7$	&	$3.86$	&	Gio88, LM97	&		&		&		&			&		\\
\hline
\multicolumn{9}{p{150mm}}{{\it References}: BP84 -- \citet{bridle84}; Ch99 -- \citet{chiaberge99}; Gio88 -- \citet{giovannini88}; GL03 -- \citet{gizani03}; LM97 -- \citet{lm97}; MKT93 -- \citet{morganti93}; NdB82 -- \citet{noordam82}.}  \\
\multicolumn{9}{p{150mm}}{$^a$log($R$) is very uncertain because of an upper limit in both the radio and optical luminosities.} \\
\multicolumn{9}{p{150mm}}{$^b$Lower limit in log($R$) because of an upper limit in the optical luminosity (see Table 4 in SSL07).} \\
\multicolumn{9}{p{150mm}}{$^c$Ch99 estimated the 5 GHz core flux density from an 8.3 GHz contour map.} \\
\end{tabular}
\end{center}
\end{table*}

\begin{table*}
\caption{Core radio luminosities and radio loudness parameters for the subsample of Palomar--Green quasars.}
\label{table: radio cores5}
\begin{center}
\begin{tabular}{lrrcclrrc}
\hline
\multicolumn{1}{c}{Source name} & \multicolumn{1}{c}{log($L_{5, \rm \: core}$)} & \multicolumn{1}{c}{log($R$)} & References &  &  Source name & \multicolumn{1}{c}{log($L_{5, \rm \: core}$)} & \multicolumn{1}{c}{log($R$)} & References \\
& \multicolumn{1}{c}{(erg s$^{-1}$)} & &  & & & \multicolumn{1}{c}{(erg s$^{-1}$)} &  &  \\
\hline
PG 0007$+$106	&	$41.1$	&	$1.98$	&	Ke89, MRS93	&		&        PG 1259$+$593	&	$<39.9$	&	$<-1.00$	&	MRS93 \\	
PG 0026$+$129	&	$<38.9$	&	$<-1.21$	&	Ke89, MRS93	&		& PG 1307$+$085	&	$<39.0$	&	$<-1.07$	&	Ke89, MRS93	\\			
PG 0050$+$124	&	$38.9$	&	$-0.67$	&	Ke89, MRS93	&		&	PG 1309$+$355	&	$41.3$	&	$1.16$	&	Ke89, MRS93	\\		 
PG 0052$+$251	&	$39.1$	&	$-0.87$	&	Ke89, MRS93	&		&	PG 1351$+$640	&	$40.2$	&	$0.77$	&	Ke89, MRS93	\\		
PG 0157$+$001	&	$40.3$	&	$0.20$	&	Ke89, MRS93	&		&	PG 1352$+$183	&	$<38.9$	&	$<-1.00$	&	Ke89, MRS93	\\		
PG 0804$+$761	&	$39.1$	&	$-0.61$	&	Ke89, MRS93	&		&	PG 1402$+$261	&	$<39.0$	&	$<-0.92$	&	Ke89, MRS93	\\		 
PG 0844$+$349	&	$<38.1$	&	$<-1.64$	&	Ke89, MRS93	&		&	PG 1411$+$442	&	$38.7$	&	$-0.98$	&	Ke89, MRS93	\\		
PG 0923$+$201	&	$<39.1$	&	$<-0.74$	&	Ke89, MRS93	&		&	PG 1415$+$451	&	$<38.6$	&	$<-0.96$	&	Ke89, MRS93	\\		 
PG 0923$+$129	&	$38.8$\rlap{$^a$}	&	$0.11$	&	MRS93		&	& PG 1416$-$129	&	$39.2$	&	$-0.56$	&	Ke89, MRS93	\\		 	
PG 0953$+$414	&	$<39.5$\rlap{$^a$}	&	$<-1.06$	&	MRS93		&	& PG 1426$+$015	&	$38.9$	&	$-0.63$	&	Ke89, MRS93	\\		 	
PG 1012$+$008	&	$39.5$	&	$-0.47$	&	Ke89, MRS93	&		& PG 1440$+$356	&	$38.7$	&	$-0.75$	&	Ke89, MRS93	 \\			
PG 1049$-$005	&	$<39.7$	&	$<-0.90$	&	Ke89, MRS93	&		& PG 1444$+$407	&	$<39.1$	&	$<-1.12$	&	Ke89, MRS93	\\			
PG 1100$+$772	&	$42.0$	&	$1.49$	&	Ke89, MRS93	&		& PG 1534$+$580	&	$38.2$	&	$-0.22$	&	Ke89, MRS93	\\		 	
PG 1103$-$006	&	$42.4$	&	$1.69$	&	Ke89, MRS93	&		&	PG 1545$+$210	&	$41.5$	&	$1.22$	&	Ke89, MRS93	\\		
PG 1114$+$445	&	$<38.7$	&	$<-0.94$	&	Ke89, MRS93	&		& PG 1613$+$658	&	$39.2$	&	$-0.57$	&	Ke89, MRS93	\\			
PG 1116$+$215	&	$39.9$	&	$-0.30$	&	Ke89, MRS93	&		&	PG 1617$+$175	&	$<39.2$\rlap{$^a$}	&	$<-0.45$	&	 Ke89, MRS93	 \\		
PG 1119$+$120	&	$<37.9$	&	$<-1.37$	&	Ke89, MRS93	&		& PG 1700$+$518	&	$40.3$	&	$-0.22$	&	Ke89, MRS93	\\			
PG 1202$+$281	&	$39.3$	&	$-0.83$	&	Ke89, MRS93	&		&	PG 2130$+$099	&	$38.7$	&	$-0.80$	&	Ke89, MRS93	\\		 
PG 1211$+$143	&	$<38.8$	&	$<-0.97$	&	MRS93	&		&	PG 2209$+$184	&	$40.8$	&	$1.75$	&	MRS93	\\		
PG 1216$+$069	&	$40.8$	&	$0.27$	&	Ke89, MRS93	&		&	PG 2251$+$113	&	$41.1$	&	$0.76$	&	Ke89, MRS93	\\		 
PG 1229$+$204	&	$<38.3$	&	$<-1.13$	&	Ke89, MRS93	&		&		PG 2308$+$098	&	$42.3$	&	$1.66$	&	Ke89, MRS93	\\		
PG 1244$+$026	&	$<38.1$	&	$<-0.55$	&	Ke89, MRS93	&		&		&		&		&		\\				
\hline
\multicolumn{9}{p{160mm}}{{\it References}: Ke89 -- \citet{kellermann89}; MRS93 -- \citet{miller93}.}  \\
\multicolumn{9}{p{160mm}}{$^a$MRS93 orginally determined 5 GHz peak flux densities for the cores of these sources from maps at 1.5 GHz, assuming $\alpha = -0.7$. For consistency with the rest of the tables, we have instead adjusted these values, and the corresponding core luminosities, so that they are consistent with $\alpha = 0$. However, for PG 1617$+$175, the total flux density at 5 GHz in Ke89 is in fact less than our extrapolated upper limit. Therefore, we have instead used the total radio luminosity as an upper limit for the core luminosity in this case.} \\
\end{tabular}
\end{center}
\end{table*}

\bsp

\label{lastpage}


\begin{thebibliography}{}

\bibitem[Anderson et al.(2004)Anderson, Ulvestad \& Ho]{anderson04} Anderson J.~M., Ulvestad J.~S., Ho L.~C., 2004, \apj, 603, 42 
\bibitem[Antonucci(1985)]{antonucci85} Antonucci R.~R.~J., 1985, \apjs, 59, 499
\bibitem[Antonucci(1993)]{antonucci93} Antonucci R., 1993, \araa, 31, 473
\bibitem[Barvainis \& Antonucci(1989)]{barvainis89} Barvainis R., Antonucci R., 1989, \apjs, 70, 257
\bibitem[Barvainis et al.(2005)]{barvainis05} Barvainis R., Leh{\'a}r J., Birkinshaw M., Falcke H., Blundell K.~M., 2005, \apj, 618, 108 
\bibitem[Becker et al.(1995)Becker, White \& Helfand]{becker95} Becker R.~H., White R.~L., Helfand D.~J., 1995, \apj, 450, 559
\bibitem[Bicknell(2002)]{bicknell02} Bicknell G.~V., 2002, New Astron. Rev., 46, 365 
\bibitem[Bicknell et al.(1998)]{bicknell98} Bicknell G.~V., Dopita M.~A., Tsvetanov Z.~I., Sutherland R.~S., 1998, \apj, 495, 680 
\bibitem[Blandford \& K\"{o}nigl(1979)]{blandford79} Blandford R.~D., K\"{o}nigl A., 1979, \apj, 232, 34 
\bibitem[Blandford \& Znajek(1977)]{blandford77} Blandford R.~D., Znajek R.~L., 1977, \mnras, 179, 433
\bibitem[Blundell et al.(2003)Blundell, Beasley \& Bicknell]{blundell03} Blundell K.~M., Beasley A.~J., Bicknell G.~V., 2003, \apjl, 591, L103 
\bibitem[Bridle \& Perley(1984)]{bridle84} Bridle A.~H., Perley R.~A., 1984, \araa, 22, 319
\bibitem[Brunthaler et al.(2000)]{brunthaler00} Brunthaler A. et al., 2000, \aap, 357, L45 
\bibitem[Burgess \& Hunstead(2006)]{burgess06} Burgess A.~M., Hunstead R.~W., 2006, \aj, 131, 114
\bibitem[Calvelo et al.(2010)]{calvelo10} Calvelo D.~E. et al., 2010, \mnras, 409, 839
\bibitem[Cao \& Rawlings(2004)]{cao04} Cao X., Rawlings S., 2004, \mnras, 349, 1419 
\bibitem[Capetti et al.(2000)]{capetti00} Capetti A., de Ruiter H.~R., Fanti R., Morganti R., Parma P., Ulrich M.-H., 2000, \aap, 362, 871 
\bibitem[Capetti et al.(2002)]{capetti02} Capetti A., Celotti A., Chiaberge M., de Ruiter H.~R., Fanti R., Morganti R., Parma P., 2002, \aap, 383, 104 
\bibitem[Chiaberge et al.(1999)Chiaberge, Capetti \& Celotti]{chiaberge99} Chiaberge M., Capetti A., Celotti A., 1999, \aap, 349, 77
\bibitem[Chiaberge et al.(2000)Chiaberge, Capetti \& Celotti]{chiaberge00} Chiaberge M., Capetti A., Celotti A., 2000, \aap, 355, 873 
\bibitem[Cirasuolo et al.(2003a)]{cirasuolo03a} Cirasuolo M., Magliocchetti M., Celotti A., Danese L., 2003a, \mnras, 341, 993 
\bibitem[Cirasuolo et al.(2003b)]{cirasuolo03b} Cirasuolo M., Celotti A., Magliocchetti M., Danese L., 2003b, \mnras, 346, 447
\bibitem[Condon et al.(1981)]{condon81} Condon J.~J., Odell S.~L., Puschell J.~J., Stein W.~A., 1981, \apj, 246, 624
\bibitem[Daly(2011)]{daly11} Daly R.~A.\ 2011, \mnras, 414, 1253 
\bibitem[De Robertis et al.(1998)De Robertis, Yee \& Hayhoe]{derobertis98} De Robertis M.~M., Yee H.~K.~C., Hayhoe K., 1998, \apj, 496, 93
\bibitem[Downes et al.(1986)]{downes86} Downes A.~J.~B., Peacock J.~A., Savage A., Carrie D.~R., 1986, \mnras, 218, 31
\bibitem[Drake et al.(2003)]{drake03} Drake C.~L., McGregor P.~J., Dopita M.~A., van Breugel W.~J.~M., 2003, \aj, 126, 2237
\bibitem[Falcke et al.(1996)Falcke, Patnaik \& Sherwood]{falcke96} Falcke H., Patnaik A.~R., Sherwood W., 1996, \apjl, 473, L13 
\bibitem[Falcke et al.(2004)Falcke, K{\"o}rding \& Markoff]{falcke04} Falcke H., K{\"o}rding E., Markoff S., 2004, \aap, 414, 895
\bibitem[Fasano \& Franceschini(1987)]{fasano87} Fasano G., Franceschini A., 1987, \mnras, 225, 155
\bibitem[Feigelson et al.(1982)Feigelson, Maccacaro \& Zamorani]{feigelson82} Feigelson E.~D., Maccacaro T., Zamorani G., 1982, \apj, 255, 392
\bibitem[Fender et al.(2004)Fender, Belloni \& Gallo]{fender04} Fender R.~P., Belloni T.~M., Gallo E., 2004, \mnras, 355, 1105
\bibitem[Fender et al.(2010)Fender, Gallo \& Russell]{fender10} Fender R.~P., Gallo E., Russell D., 2010, \mnras, 406, 1425
\bibitem[Gallo et al.(2003)Gallo, Fender \& Pooley]{gallo03} Gallo E., Fender R.~P., Pooley G.~G., 2003, \mnras, 344, 60
\bibitem[Gilbert et al.(2004)]{gilbert04} Gilbert G.~M., Riley J.~M., Hardcastle M.~J., Croston J.~H., Pooley G.~G., Alexander P., 2004, \mnras, 351, 845
\bibitem[Giovannini et al.(1988)]{giovannini88} Giovannini G., Feretti L., Gregorini L., Parma P., 1988, \aap, 199, 73
\bibitem[Giovannini et al.(2001)]{giovannini01} Giovannini G., Cotton W.~D., Feretti L., Lara L., Venturi T., 2001, \apj, 552, 508
\bibitem[Gizani \& Leahy(2003)]{gizani03} Gizani N.~A.~B., Leahy J.~P., 2003, \mnras, 342, 399 
\bibitem[Gopal-Krishna et al.(2008)Gopal-Krishna, Mangalam \& Wiita]{gopal08} Gopal-Krishna, Mangalam A., Wiita P.~J., 2008, \apjl, 680, L13
\bibitem[Gregory \& Condon(1991)]{gregory91} Gregory P.~C., Condon J.~J., 1991, \apjs, 75, 1011
\bibitem[G{\"u}ltekin et al.(2009)]{gultekin09} G{\"u}ltekin K., Cackett E.~M., Miller J.~M., Di Matteo T., Markoff S., Richstone D.~O., 2009, \apj, 706, 404 
\bibitem[Hardcastle \& Worrall(2000)]{hardcastle00} Hardcastle M.~J., Worrall D.~M., 2000, \mnras, 314, 359 
\bibitem[Hardcastle et al.(2009)Hardcastle, Evans \& Croston]{hardcastle09} Hardcastle M.~J., Evans D.~A., Croston J.~H., 2009, \mnras, 396, 1929 
\bibitem[Healey et al.(2007)]{healey07} Healey S.~E., Romani R.~W., Taylor G.~B., Sadler E.~M., Ricci R., Murphy T., Ulvestad J.~S., Winn J.~N., 2007, \apjs, 171, 61
\bibitem[Heinz(2002)]{heinz02} Heinz S., 2002, \aap, 388, L40
\bibitem[Heinz \& Sunyaev(2003)]{heinz03} Heinz S., Sunyaev R.~A., 2003, \mnras, 343, L59
\bibitem[Ho(2002)]{ho02} Ho L.~C., 2002, \apj, 564, 120
\bibitem[Ho \& Peng(2001)]{ho01} Ho L.~C., Peng C.~Y., 2001, \apj, 555, 650
\bibitem[Ho \& Ulvestad(2001)]{houl01} Ho L.~C., Ulvestad J.~S., 2001, \apjs, 133, 77
\bibitem[Jones et al.(1994)Jones, McAdam \& Reynolds]{jones94} Jones P.~A., McAdam W.~B., Reynolds J.~E., 1994, \mnras, 268, 602
\bibitem[Kapahi(1995)]{kapahi95} Kapahi V.~K., 1995, \jaa, 16, 1
\bibitem[Kellermann et al.(1989)]{kellermann89} Kellermann K.~I., Sramek R., Schmidt M., Shaffer D.~B., Green R., 1989, \aj, 98, 1195
\bibitem[K{\"o}rding et al.(2006)K{\"o}rding, Fender \& Migliari]{koerding06b} K{\"o}rding E.~G., Fender R.~P., Migliari S., 2006, \mnras, 369, 1451 
\bibitem[K{\"o}rding et al.(2006)K{\"o}rding, Falcke \& Corbel]{koerding06} K{\"o}rding E., Falcke H., Corbel S., 2006, \aap, 456, 439
\bibitem[Kukula et al.(1995)]{kukula95} Kukula M.~J., Pedlar A., Baum S.~A., O'Dea C.~P., 1995, \mnras, 276, 1262
\bibitem[Ivezi{\'c} et al.(2002)]{ivezic02} Ivezi{\'c} {\v Z}. et al., 2002, \aj, 124, 2364
\bibitem[Lagos et al.(2009)Lagos, Padilla \& Cora]{lagos09} Lagos C.~D.~P., Padilla N.~D., Cora S.~A., 2009, \mnras, 395, 625
\bibitem[Lal et al.(2011)Lal, Shastri \& Gabuzda]{lal11} Lal D.~V., Shastri P., Gabuzda D.~C., 2011, \apj, 731, 68 
\bibitem[Laor(2003)]{laor03} Laor A., 2003, arXiv:astro-ph/0312417
\bibitem[Landt et al.(2002)Landt, Padovani \& Giommi]{landt02} Landt H., Padovani P., Giommi P., 2002, \mnras, 336, 945 
\bibitem[Laurent-Muehleisen et al.(1997)]{lm97} Laurent-Muehleisen S.~A., Kollgaard R.~I., Ryan P.~J., Feigelson E.~D., Brinkmann W., Siebert J., 1997, \aaps, 122, 235
\bibitem[Lister et al.(1994)Lister, Gower \& Hutchings]{lister94} Lister M.~L., Gower A.~C., Hutchings J.~B., 1994, \aj, 108, 821
\bibitem[Maccarone et al.(2003)Maccarone, Gallo \& Fender]{maccarone03} Maccarone T.~J., Gallo E., Fender R., 2003, \mnras, 345, L19
\bibitem[McLure \& Dunlop(2002)]{mclure02} McLure R.~J., Dunlop J.~S., 2002, \mnras, 331, 795 
\bibitem[McNamara et al.(2011)McNamara, Rohanizadegan \& Nulsen]{mcnamara11} McNamara B.~R., Rohanizadegan M., Nulsen P.~E.~J., 2011, \apj, 727, 39 
\bibitem[McKinney(2005)]{mckinney05} McKinney J.~C., 2005, \apjl, 630, L5
\bibitem[Mart{\'{\i}}nez-Sansigre \& Rawlings(2011)]{sansigre11} Mart{\'{\i}}nez-Sansigre A., Rawlings S.\ 2011, \mnras, 414, 1937 
\bibitem[Meier(2001)]{meier01} Meier D.~L., 2001, \apjl, 548, L9 
\bibitem[Merloni et al.(2003)Merloni, Heinz \& Di Matteo]{merloni03} Merloni A., Heinz S., Di Matteo T., 2003, \mnras, 345, 1057
\bibitem[Middelberg et al.(2004)]{middleberg04} Middelberg E. et al., 2004, \aap, 417, 925 
\bibitem[Migliari \& Fender(2006)]{migliari06} Migliari S., Fender R.~P., 2006, \mnras, 366, 79
\bibitem[Miller et al.(1990)Miller, Peacock \& Mead]{miller90} Miller L., Peacock J.~A., Mead A.~R.~G., 1990, \mnras, 244, 207
\bibitem[Miller et al.(1993)Miller, Rawlings \& Saunders]{miller93} Miller P., Rawlings S., Saunders R., 1993, \mnras, 263, 425
\bibitem[Morganti et al.(1993)Morganti, Killeen \& Tadhunter]{morganti93} Morganti R., Killeen N.~E.~B., Tadhunter C.~N., 1993, \mnras, 263, 1023
\bibitem[Mullin \& Hardcastle(2009)]{mullin09} Mullin L.~M., Hardcastle M.~J., 2009, \mnras, 398, 1989 
\bibitem[Murphy et al.(2010)]{murphy10} Murphy T. et al., 2010, \mnras, 402, 2403
\bibitem[Nagar et al.(2005)Nagar, Falcke \& Wilson]{nagar05} Nagar N.~M., Falcke H., Wilson A.~S., 2005, \aap, 435, 521 
\bibitem[Noordam \& de Bruyn(1982)]{noordam82} Noordam J.~E., de Bruyn A.~G., 1982, \nat, 299, 597
\bibitem[Orr \& Browne(1982)]{orr82} Orr M.~J.~L., Browne I.~W.~A., 1982, \mnras, 200, 1067
\bibitem[Perley(1982)]{perley82} Perley R.~A., 1982, \aj, 87, 859
\bibitem[Peterson(1997)]{peterson97} Peterson B.~M., 1997, An Introduction to Active Galactic Nuclei. Cambridge Univ. Press, Cambridge
\bibitem[Press et al.(1992)]{press92} Press W.~H., Teukolsky S.~A., Vetterling W.~T., Flannery B.~P., 1992, Numerical Recipes in C. The Art of Scientific Computing, 2nd ed. Cambridge Univ. Press, Cambridge
\bibitem[Prestage \& Peacock(1988)]{prestage88} Prestage R.~M., Peacock J.~A., 1988, \mnras, 230, 131
\bibitem[Rafter et al.(2011)Rafter, Crenshaw \& Wiita]{rafter11} Rafter S.~E., Crenshaw D.~M., Wiita P.~J., 2011, \aj, 141, 85 
\bibitem[Reid et al.(1999)Reid, Kronberg \& Perley]{reid99} Reid R.~I., Kronberg P.~P., Perley R.~A., 1999, \apjs, 124, 285
\bibitem[Roy et al.(1994)]{roy94} Roy A.~L., Norris R.~P., Kesteven M.~J., Troup E.~R., Reynolds J.~E., 1994, \apj, 432, 496
\bibitem[Sanghera et al.(1995)]{sanghera95} Sanghera H.~S., Saikia D.~J., Luedke E., Spencer R.~E., Foulsham P.~A., Akujor C.~E., Tzioumis A.~K., 1995, \aap, 295, 629
\bibitem[Schmitt(1985)]{schmitt85} Schmitt J.~H.~M.~M., 1985, \apj, 293, 178 
\bibitem[Sikora(2009)]{sikora09} Sikora M., 2009, Astron. Nachr., 330, 291
\bibitem[Sikora et al.(2007)Sikora, Stawarz \& Lasota]{sikora07} Sikora M., Stawarz {\L}., Lasota J.-P., 2007, \apj, 658, 815 (SSL07)
\bibitem[Singal et al.(2011)]{singal11} Singal J., Petrosian V., Lawrence A., Stawarz L., 2011, arXiv:1101.2930 
\bibitem[Singh \& Westergaard(1992)]{singh92} Singh K.~P., Westergaard N.~J., 1992, \aap, 264, 489
\bibitem[Sramek \& Weedman(1980)]{sramek80} Sramek R.~A., Weedman D.~W., 1980, \apj, 238, 435
\bibitem[Stanghellini et al.(1997)]{stanghellini97} Stanghellini C., Bondi M., Dallacasa D., O'Dea C.~P., Baum S.~A., Fanti R., Fanti C., 1997, \aap, 318, 376
\bibitem[Strittmatter et al.(1980)]{strittmatter80} Strittmatter P.~A., Hill P., Pauliny-Toth I.~I.~K., Steppe H., Witzel A., 1980, \aap, 88, L12
\bibitem[Terashima \& Wilson(2003)]{terashima03} Terashima Y., Wilson A.~S., 2003, \apj, 583, 145
\bibitem[Tchekhovskoy et al.(2010)Tchekhovskoy, Narayan \& McKinney]{tchekhovskoy10} Tchekhovskoy A., Narayan R., McKinney J.~C., 2010, \apj, 711, 50
\bibitem[Ulvestad \& Wilson(1984a)]{ulvestad84a} Ulvestad J.~S., Wilson A.~S., 1984a, \apj, 278, 544
\bibitem[Ulvestad \& Wilson(1984b)]{ulvestad84b} Ulvestad J.~S., Wilson A.~S., 1984b, \apj, 285, 439
\bibitem[Ulvestad et al.(1999)]{ulvestad99} Ulvestad J.~S., Wrobel J.~M., Roy A.~L., Wilson A.~S., Falcke H., Krichbaum T.~P., 1999, \apjl, 517, L81 
\bibitem[Unger et al.(1987)]{unger87} Unger S.~W., Lawrence A., Wilson A.~S., Elvis M., Wright A.~E., 1987, \mnras, 228, 521
\bibitem[Vestergaard(2002)]{vestergaard02} Vestergaard M., 2002, \apj, 571, 733 
\bibitem[Volonteri et al.(2007)Volonteri, Sikora \& Lasota]{volonteri07} Volonteri M., Sikora M., Lasota J.-P., 2007, \apj, 667, 704
\bibitem[Wang \& Kaiser(2008)]{wang08} Wang Y., Kaiser C.~R., 2008, \mnras, 388, 677
\bibitem[White et al.(2000)]{white00} White R.~L. et al., 2000, \apjs, 126, 133
\bibitem[Wills \& Browne(1986)]{wills86} Wills B.~J., Browne I.~W.~A., 1986, \apj, 302, 56
\bibitem[Wilson \& Tsvetanov(1994)]{wilson94} Wilson A.~S., Tsvetanov Z.~I., 1994, \aj, 107, 1227
\bibitem[Winn et al.(2003)Winn, Patnaik \& Wrobel]{winn03} Winn J.~N., Patnaik A.~R., Wrobel J.~M., 2003, \apjs, 145, 83
\bibitem[Woo \& Urry(2002)]{woo02} Woo J.-H., Urry C.~M., 2002, \apj, 579, 530 
\bibitem[Wright(2006)]{wright06} Wright E.~L., 2006, \pasp, 118, 1711
\bibitem[Xu et al.(1999)Xu, Livio \& Baum]{xu99} Xu C., Livio M., Baum S., 1999, \aj, 118, 1169
\bibitem[York et al.(2000)]{york00} York D.~G. et al., 2000, \aj, 120, 1579
\bibitem[Zamfir et al.(2008)Zamfir, Sulentic \& Marziani]{zamfir08} Zamfir S., Sulentic J.~W., Marziani P., 2008, \mnras, 387, 856
\bibitem[Zirbel \& Baum(1995)]{zirbel95} Zirbel E.~L., Baum S.~A., 1995, \apj, 448, 521
\end{thebibliography}
\end{document}